\documentclass[a4paper,11pt]{article}
\pdfoutput=1
\usepackage{jheppub}
\usepackage{amsfonts,amssymb,amsmath}
\usepackage[normalem]{ulem}
\usepackage{color}
\usepackage{epstopdf}
\usepackage{tensor}
\usepackage{enumerate}
\usepackage{pdflscape}
\usepackage{float}

\newcommand{\be}{\begin{equation}}
\newcommand{\ee}{\end{equation}}
\newcommand{\ba}{\begin{eqnarray}}
\newcommand{\ea}{\end{eqnarray}}
\newcommand{\beal}{\begin{aligned}}
\newcommand{\eeal}{\end{aligned}}
\newcommand{\nn}{\nonumber}

\title{Criticality for charged black branes}

\author{Robie A. Hennigar}

\affiliation{Department of Physics and Astronomy, University of Waterloo,
\\
 200 University Avenue West, Waterloo, Ontario, N2L 3G1, Canada}

\emailAdd{rhennigar@uwaterloo.ca}

\abstract
{
We show that the inclusion of higher curvature terms in the gravitational action can lead to phase transitions and critical behaviour for charged black branes.  The higher curvature terms considered here belong to the recently constructed generalized quasi-topological class [ArXiv: 1703.01631], which possess a number of interesting properties, such as being ghost-free on constant curvature backgrounds and non-trivial in four dimensions.  We show that critical behaviour is a generic feature of the black branes in all dimensions $d \ge 4$, and contextualize the results with a review of the properties of black branes in Lovelock and quasi-topological gravity, where critical behaviour is not possible.  These results may have interesting implications for the CFTs dual to this class of theories.}

\begin{document}

\maketitle

\section{Introduction}

That black holes have thermodynamic properties remains one of the most significant insights toward a quantum theory of gravity made in the last forty years, revealing evidence for the microscopic structure of gravity. In the case of anti de Sitter (AdS) black holes, through the conjectured anti de Sitter/conformal field theory (CFT) correspondence, it is possible to deduce properties of the holographically dual CFTs via the study of AdS black holes~\cite{Hawking:1982dh, Witten:1998zw}, even in the limit of strong coupling. A particularly interesting case is that of planar AdS black holes---\textit{black branes}---whose thermodynamics and hydrodynamics correspond to those of a finite temperature quantum field theory with translation invariance in one less dimension. 

A rich arena for the study of black objects is provided by higher curvature gravity.  In these theories, the Einstein-Hilbert action is modified by the inclusion of additional terms non-linear in the curvature.  The motivation for these corrections comes in many forms, but is primarily due to the expectation that they will appear in any ultraviolet completion of general relativity.  At the quantum level, including quadratic terms in the action can lead to a predicative and renormalizable theory of quantum gravity where all scales are generated dynamically~\cite{Stelle:1976gc, Salvio:2014soa}, various higher curvature terms appear in the low energy limit of string theory~\cite{Zwiebach:1985uq, Metsaev:1986yb, Gross:1986mw, Myers:1987qx}, and from the perspective of holography, higher curvature terms correspond to corrections from a finite number of colors, $1/N_c$, and therefore allow one to make contact with a broader class of CFTs via holography~\cite{Myers:2010jv, Son:2007vk,Buchel:2008sh,Kats:2007mq, Myers:2008yi, Buchel:2008vz, Casadio:2016zhu}.

The focus of this study will be an investigation of charged black brane solutions in higher curvature gravity from the perspective of \textit{black hole chemistry}. Here the cosmological constant is considered as a thermodynamic variable~\cite{Henneaux:1985tv, Creighton:1995au, Caldarelli:1999xj}, interpreted as pressure in the first law of black hole mechanics~\cite{Kastor:2009wy, KastorEtal:2010}. In this framework,  there is a physical analogy between the charged AdS black hole and the van der Waals fluid, with the analog of the liquid/gas transition being a small/large black hole phase transition~\cite{Kubiznak:2012wp}.  Studies of black holes with spherical or hyperbolic horizons in higher curvature gravities have revealed a variety of nontrivial behaviour, including novel universality classes associated with isolated critical points~\cite{Dolan:2014vba, Hennigar:2015esa, Dykaar:2017mba}, triple points and (multiple) reentrant phase transitions~\cite{Frassino:2014pha, Hennigar:2015esa}, and superfluid-like phase transitions~\cite{  Hennigar:2016xwd, Dykaar:2017mba, Kubiznak:2016qmn}, with similar results found in numerous other studies~\cite{Wei:2012ui, Cai:2013qga, Xu:2013zea, Mo:2014qsa, Wei:2014hba, Mo:2014mba, Zou:2014mha, Belhaj:2014tga, Xu:2014tja, Frassino:2014pha, Dolan:2014vba, Belhaj:2014eha, Sherkatghanad:2014hda, Hendi:2015cka, Hendi:2015oqa, Hennigar:2015esa, Hendi:2015psa, Nie:2015zia, Hendi:2015pda, Hendi:2015soe, Johnson:2015ekr, Hendi:2016njy, Zeng:2016aly, Hennigar:2016gkm, Puletti:2017gym}.  These results provide hints toward the rich structure that emerges for gravity when quantum effects are taken into account, and also to the phase structure of the holographically dual gauge theories.  

Within higher curvature gravity, the holographic study of black brane geometries has led to numerous interesting results---for example, the inclusion of quadratic terms has been shown to lead to violations of  the Kovtun--Son--Starinets (KSS) viscosity/entropy ratio bound~\cite{Kovtun:2004de, Kats:2007mq, Brigante:2007nu, Buchel:2008vz}.  However, while AdS black objects with spherical or hyperbolic horizon topologies have been found to present a variety of interesting critical behaviour and phase structure, black branes have thus far been found to be devoid of any such behaviour.   This is linked to the fact that, in many commonly studied higher curvature theories, black brane solutions are not modified from their Einstein gravity form---that is, they possess universal properties.

Here we will show that the inclusion of certain higher curvature terms in the gravitational action can lead to non-trivial thermodynamic behaviour for electrically charged  black branes.  The theories considered belong to the \textit{generalized quasi-topological gravity} (GQG) class~\cite{Hennigar:2017ego}, which describe theories that, under the restriction to spherically (planar or hyperbolic) symmetric metrics, admit a single independent field equation in vacuum (and also for suitable matter, such as a Maxwell field)~\cite{Jacobson:2007tj, Giambo:2002wr, Salgado:2003ub}.\footnote{The construction of these theories was motivated by the recently discovered \textit{Einsteinian cubic gravity}~\cite{Bueno:2016xff}, which was found to be the unique cubic theory of gravity with a single independent field equation for a four dimensional spherically symmetric metric~\cite{Hennigar:2016gkm, Bueno:2016lrh}.  Einsteinian cubic gravity was constructed in~\cite{Bueno:2016xff} (see also~\cite{Bueno:2016ypa}) with the aim of presenting the most general cubic theory which is: (a) ghost-free on a maximally symmetric background and (b) has the coefficients in the action independent of spacetime dimension. That the theory (in four dimensions) possesses a single independent field equation was an unexpected result, and the generalized quasi-topological class of theories generalizes this property (which holds also for Lovelock~\cite{Lovelock:1971yv} and quasi-topological gravity~\cite{Myers:2010ru, Oliva:2010eb}) to all dimensions.}  Subsequent work has revealed that these theories possess a number of interesting properties.  In~\cite{Hennigar:2017ego}, it was conjectured that any theory for which the \textit{most general} static, spherically symmetric solution can be written in the form,
\be 
ds^2 = -f(r) dt^2 + \frac{dr^2}{f(r)} + r^2 d\Sigma^2_{k, (d-2)}
\ee
(i.e. with a single metric function) does not propagate ghosts on a maximally symmetric background.  This result was later proven to be correct~\cite{Bueno:2017sui}, indicating that the entire class of GQG theories are free from ghosts.   The theories are non-trivial in four dimensions and provide the only non-trivial examples of ghost-free higher curvature theories active in four dimensions~\cite{Hennigar:2017ego, Bueno:2017sui, Ahmed:2017jod}.  Despite third order equations of motion, the black hole solutions possess no higher derivative hair~\cite{Goldstein:2017rxn} (the solutions are characterized only by their mass), and the thermodynamic properties can be studied exactly, despite the lack of an exact solution~\cite{Hennigar:2017ego, Bueno:2017sui, Ahmed:2017jod}.  Unlike  Lovelock~\cite{Lovelock:1971yv} and quasi-topological gravity~\cite{Myers:2010ru, Oliva:2010eb}, the properties of black branes are modified by the presence  of these terms, a fact which may lead to interesting consequences for the dual CFTs~\cite{Bueno:2017sui, Ahmed:2017jod}.  More recently, it has been shown, in four dimensions, that these higher curvature terms generically result in the specific heat of asymptotically flat black holes becoming positive below a certain critical size, signalling the onset of thermodynamic stability for small black holes~\cite{Bueno:2017qce}.  

This paper is organized as follows.  We begin in section~\ref{sec:LL_QT} with a review of the properties of black branes in Lovelock and quasi-topological gravity, emphasizing that black branes in these theories are exactly the same as the corresponding solutions to Einstein gravity.  We then, in section~\ref{sec:charged_BB}, construct charged black brane solutions in the cubic and quartic versions of generalized quasi-topological gravity, exploring the solutions via series approximations asymptotically, near the horizon, and near the origin; numerical calculations allow for the series approximations to be joined in the intermediate regimes.  In section~\ref{sec:higher_d_thermo} we consider the critical behaviour of the black branes in arbitrary dimensions larger than five.  After enforcing various physicality constraints, we show that critical points and phase transitions occur, with novel structure observed for the free energy---including three branches of solutions with positive specific heat.  Lastly, in section~\ref{sec:four_d}, we consider seperately the case of four dimensions, where the general form of the generalized quasi-topological class is known to arbitrary order in the curvature.  Here we show similar results hold, including phase transitions and critical points characterized by mean field theory critical exponents.

\section{Black branes in higher curvature gravities}
\label{sec:LL_QT}

Numerous interesting results have been obtained through the study of black branes in higher curvature gravity.  However, in many of the models that are frequently employed for computational ease---such as Lovelock and quasi-topological gravity---the thermodynamic properties of black branes are, in a sense, universal~\cite{Cadoni:2016hhd}.  Here we briefly review this to contextualize the results that will be obtained in the following sections for the generalized quasi-topological theories.  For brevity, we will skip over many of the explicit calculations, referencing work where the relevant details can be found.

We will consider an action comprised of Lovelock and quasi-topological gravity with terms up to third order in curvature.  The essential results extend to arbitrary order in the curvature, but restricting attention to third order is sufficient for highlighting the salient details.  The action is given by,
\begin{align}
\mathcal{I} =& \frac{1}{16 \pi G} \int d^d x \sqrt{-g} \bigg[\frac{(d-1)(d-2)}{\ell^2} + R + \frac{\alpha_2}{(d-3)(d-4)} \mathcal{X}_4 - \frac{1}{4} F_{ab}F^{ab}
\nn\\
&+ \frac{\alpha_3}{(d-3)(d-4)(d-5)(d-6)} \mathcal{X}_6 + \frac{8(2d-3) \rho}{(d-3)(d-6)(3d^2 - 15d + 16)}  \mathcal{Z}_d  \bigg] \, .
\end{align}
In the above action, the terms $\mathcal{X}_4$ and $\mathcal{X}_6$ refer to the four- and six-dimensional Euler densities, which correspond to the standard Gauss-Bonnet and cubic Lovelock terms.  The term $\mathcal{Z}_d$ is the cubic quasi-topological term~\cite{Myers:2010ru, Oliva:2010eb}.  The explicit form of the Lagrangians are listed in appendix~\ref{sec:lagrangians} for brevity.

To consider black brane solutions of the theory, we substitute the metric ansatz,
\be 
ds^2 = -N(r)^2 f(r) dt^2 + \frac{dr^2}{f(r)} + r^2\sum_{i = 1}^{d-2} dx_i^2 
\ee
into the action and vary with respect to $N(r)$ and $f(r)$ to obtain the two independent (gravitational) field equations.  The field equation following from the variation of $f(r)$ can be solved by setting $N(r) = const.$ (here we will choose $N(r) = N_0$), while the remaining field equation reduces to a polynomial in $f(r)$:
\be 
r^{d-1} \left[\frac{1}{\ell^2} - \frac{f}{r^2} + \alpha_2 \frac{f^2}{r^4} + (\rho - \alpha_3) \frac{f^3}{r^6} \right] = \omega^{d-3} - \frac{q^2}{r^{d-3}} \, .
\ee
In the above, $\omega$ is an integration constant with units of length,  and $q$ is an integration constant from the Maxwell equations and is related to the electric charge (see, e.g. \cite{Myers:2010ru, Hennigar:2015esa}).

To study the thermodynamic properties of the black branes, we proceed by considering the metric function expanded near the horizon.  In this limit we write,
\be 
f(r) = \frac{4 \pi T}{N_0}(r-r_+) + a_2(r-r_+)^2 + a_3 (r-r_+)^3 + \cdots
\ee
Substituting this expression for $f(r)$ into the field equations allows one to determine the constant $\omega$ and the temperature as,
\begin{align}
\omega^{d-3} &= \frac{r_+^{d-1}}{\ell^2} + \frac{q^2}{r_+^{d-3}} 
\, ,
\nn\\
T &= \frac{(d-1) N_0 r_+}{4 \pi \ell^2} - \frac{(d-3) N_0 q^2}{4 \pi r_+^{2d - 5}} \, .
\end{align}
The key aspect of the above result is that the higher curvature terms \textit{have no effect on these thermodynamic parameters}, except possibly through dependence implicit in the arbitrary constant $N_0$.  The above expressions for $\omega$ and the temperature are exactly the same as the corresponding expressions in Einstein-AdS-Maxwell gravity.  

Going further, we can use the expression for the temperature above to write down the equation of state~\cite{Hennigar:2015esa},
\be 
P = \frac{T}{v} + \frac{16^{d-3} (d-3) N_0^{2d-4} }{\pi (d-2)^{2d-5}  } \frac{q^2}{v^{2d-4}} \, .
\ee 
Again, the equation of state for the black branes is the same as what would be obtained in Einstein-AdS-Maxwell gravity, and the higher curvature terms make no contribution (except perhaps entering through the arbitrary constant $N_0$).  This fact remains unchanged in Lovelock and quasi-topological gravity to any order in the curvature.  As a consequence, there is no critical behaviour for electrically charged black branes in these theories in any spacetime dimension.

The thermodynamic properties of black branes in these theories have been considered in a number of studies~\cite{Myers:2010ru, Hennigar:2015esa, Ahmed:2017jod, Frassino:2014pha}.  The basic thermodynamic quantities are (neglecting the infinite volume of the transverse space and reporting densities),
\begin{align}
M &= \frac{(d-2)N_0 \omega^{d-3}}{16 \pi }\, , \quad s = \frac{r_+^4}{4} \, , \quad Q = \frac{\sqrt{2(d-2)(d-3)} N_0}{16 \pi }q 
\nn\\
\Phi &= \sqrt{\frac{2(d-2)}{(d-3)}} \frac{q}{r^{d-3}} \, , \quad P = \frac{(d-1)(d-2)}{16 \pi \ell^2} \, , \quad V = \frac{N_0 r_+^{d-1}}{d-1} \,,
\end{align}
and satisfy the (extended) first law,
\be 
dM = T ds + V dP + \Phi dQ \, ,
\ee
and the Smarr formula which follows from scaling,
\be 
(d-3) M = (d-2) Ts - 2 P V + (d-3) \Phi Q \, .
\ee
Alternatively, since
\be 
(d-1) PV = (d-2) Ts \, ,
\ee
a feature of planar black holes that was noted in~\cite{Brenna:2015pqa}, these quantities also satisfy,
\be 
M = \frac{d-2}{d-1} \left[Ts + \Phi Q \right]
\ee
as expected for a CFT in $d-1$ dimensions with a chemical potential.

It is noteworthy that the entropy of the black branes are unaffected by the presence of the higher curvature terms.  These terms modify the entropy of spherical and topological black holes~\cite{Frassino:2014pha}, but the entropy of the black branes receives no correction.  Additionally, in the black hole chemistry framework, the potentials which are conjugate to the higher curvature couplings also vanish for black branes in these theories~\cite{Frassino:2014pha}, which simplifies their thermodynamic description.

The material reviewed above highlights the fact that the thermodynamic properties of charged black branes are universal in Lovelock and quasi-topological gravity---that is, the properties of the black branes are unaffected by the higher curvature terms, except possibly through a choice of lapse function which depends on these couplings.  While the above discussion was presented for the concrete case of third order Lovelock and quasi-topological gravity, the results go through exactly the same in any higher order version of these theories. As a result, there is no critical behaviour for black branes at any order of Lovelock or quasi-topological gravity, or in any dimension.

\section{Charged black brane solutions in cubic and quartic GQG}
\label{sec:charged_BB}

The goal of this section is to demonstrate that, within generalized quasi-topological gravity, charged black branes can have critical points and can undergo phase transitions.  In light of this goal, and in view of the discussion in the previous section, here we will consider only Einstein gravity supplemented by the cubic and quartic generalized quasi-topological terms.  This choice is made for simplicity in the analysis, since we know that Lovelock or quasi-topological terms will have no effect on the thermodynamic relations.  We consider a theory of gravity in $d$-dimensions given by the following action,
\begin{align}
\label{eqn:cubic_action}
\mathcal{I} =& \frac{1}{16 \pi G} \int d^dx \sqrt{-g} \bigg[\frac{(d-1)(d-2)}{\ell^2} + R + \hat{\lambda}\mathcal{S}_{3,d} + \hat{\mu} \mathcal{S}_{4,d}   - \frac{1}{4} F_{ab} F^{ab} \bigg] \, .
\end{align}
In addition to the Einstein-Hilbert terms, we have the cubic generalized quasi-topological term $\mathcal{S}_{3,d}$,
\ba\label{Sd}
\mathcal{S}_{3, d} &=&
14 R_{a}{}^{e}{}_{c}{}^{f} R^{abcd} R_{bedf}+ 2 R^{ab} R_{a}{}^{cde} R_{bcde}- \frac{4 (66 - 35 d + 2 d^2) }{3 (d-2) (2 d-1)} R_{a}{}^{c} R^{ab} R_{bc}
\nn\\
&&-  \frac{2 (-30 + 9 d + 4 d^2) }{(d-2) (2 d-1)} R^{ab} R^{cd} R_{acbd} 
-  \frac{(38 - 29 d + 4 d^2)}{4 (d -2) (2 d  - 1)} R R_{abcd} R^{abcd}  
\nn\\
&&+ \frac{(34 - 21 d + 4 d^2) }{(d-2) ( 2 d - 1)} R_{ab} R^{ab} R -  \frac{(30 - 13 d + 4 d^2)}{12 (d-2) (2 d - 1)}  R^3 
\ea
and the quartic generalized quasi-topological term, $\mathcal{S}_{4,d}$.  The Lagrangian density for the quartic term is rather complicated and we have included it in the appendix (see also~\cite{Ahmed:2017jod}). 
In the action, the coupling constants with hats are given by,
\begin{align}
\hat{\lambda} &= - \frac{12(2d-1)(d-2)}{d-3} \lambda \, ,
\nn\\
\hat{\mu} &= - {\frac { d\left( 3{d}^{3}-27{d}^{2}+73\,d-57 \right) }{8
 \left( {d}^{5}-14{d}^{4}+79{d}^{3}-224{d}^{2}+316\,d-170
 \right)   }} \mu
\end{align} 
where $\lambda$ and $\mu$ are arbitrary coupling constants, and the prefactors have been chosen to simplify the resulting field equations.  Due to the design of this theory, it allows for planar black hole solutions of the form,
\be 
ds^2 = -N(r)^2 f(r) dt^2 + \frac{dr^2}{f(r)} + r^2\sum_{i = 1}^{d-2} dx_i^2 
\ee
with $N(r) = {\rm constant}$.\footnote{This property holds for spherical, planar and hyperbolic solutions of the theory.  In fact, this is the defining feature of the generalized quasi-topological class~\cite{Hennigar:2017ego}.}  Here we shall take $N = 1$ for simplicity.\footnote{A common convention is to take $N = 1/\sqrt{f_\infty}$,  amounts to normalizing the speed of light on the boundary (or in the dual CFT) to be $c = 1$.  However, there is no loss of generality in setting $N = 1$, due to the time reparameterization invariance of the metric.}

At large $r$, the metric approaches AdS space, with the metric function behaving as,
\be 
f \approx f_\infty \frac{r^2}{\ell^2} \quad \text{as $r \to \infty$}
\ee
where $f_\infty$ describes the asymptotic behaviour of the metric function, and solves the polynomial equation,
\be\label{eqn:asymp_f} 
1 - f_\infty - \frac{\lambda}{\ell^4} (d-6)(4d^4 - 49 d^3 + 291 d^2 - 514 d + 184) f_\infty^3 + \frac{\mu}{\ell^6} \frac{(d-8)}{3} f_\infty^4  = 0 \, .
\ee
It is a general consequence of adding higher curvature terms to the action that the cosmological constant does not describe the curvature of the asymptotic AdS region.  Also, since we are considering only asymptotically AdS spaces, $f_\infty$ must be a positive quantity.  The derivative of this equation (with respect to $f_\infty$) gives the prefactor appearing in the linearized equations of motion, and thus determines the sign of the effective Newton's constant in this theory.  Demanding that this constant has the same sign as it does in Einstein gravity requires that the function,
\be\label{eqn:asymp_der} 
P(f_\infty) =  1 +  3\frac{\lambda}{\ell^4} (d-6)(4d^4 - 49 d^3 + 291 d^2 - 514 d + 184) f_\infty^2 - \frac{\mu}{\ell^6} \frac{4(d-8)}{3} f_\infty^3 
\ee 
is positive.  This ensures that gravity couples to matter in the correct way, and is equivalent to demanding that the graviton is not a ghost. 

We have included a Maxwell field $F_{ab} = \partial_a A_b - \partial_b A_a$, and parameterize the electromagnetic one-form as,
\be 
A = q E(r) dt \, .
\ee
Solving the Maxwell equations, we find that the unknown function can be written as,
\be 
E(r) = \sqrt{\frac{2(d-2)}{d-3}} \frac{1}{r^{d-3}} \, ,
\ee
where the dimension-dependent prefactor has been chosen for later convenience. 

The field equation determining the metric function, $f$ takes the form $(d-2)F' = 0$ where
\begin{align}
F &= r^{d-1} \left[ \frac{1}{\ell^2} - \frac{f}{r^2}  \right] - \lambda F_{\mathcal{S}_{3,d}} + \mu F_{\mathcal{S}_{4,d}}  +  q^2 r^{3-d} \, .
\end{align} 
The terms $F_{\mathcal{S}_{3,d}}$ and $F_{\mathcal{S}_{4,d}}$ are the contributions to the field equations from the cubic and quartic generalized quasi-topological terms, and their explicit form for planar transverse geometries are,
\begin{align}
F_{\mathcal{S}_{3, d}} &=    \left( d-4 \right) f ^{3}  \bigg[  4 {d}^{4} - 57 {d}^{3}+  1044  {d}^{2} - 1248  d + 1956 \bigg]  {r}^
{d-7} 
\nn\\
& +96\left({d}^{2} +  5d-15 \right)  \bigg[  r^{d-5} f f'' \left(f - r \frac{f'}{2} \right) + \frac{f'^3}{6} r^{d-4}  
- \frac{1}{4} (d-4)fr^{d-5} + (d-5)f^2f'r^{d-6} \bigg]  \, ,
\nn\\
F_{\mathcal{S}_{4,d}} &=  f^2 \left[    \left( d-4
 \right) f  f'' -  {f'}^{2} \left({d}^{2}-\frac{23}{2} d + 32
 \right)  \right] {r}^{d-7}
 \nn\\
 & - 
  2 f f' f'' \left( f    \left( d-5 \right) {r}^{d-6} - 
\frac{f'}{8} \left( 3d- 16 \right) {r}^{d-5} \right)   
+f^3 f' \left( d-4 \right)  \left( d-7 \right) {r}^{d-8}
\nn\\
&+\frac{f'^3}{12}\left( 3d- 16 \right) \bigg[   \left( d-5 \right) f {r}^{d-6} 
- 3 \frac{f'}{4}{r}^{d-5} \bigg] \, .
\end{align}
Integrating this equation, we write
\be 
F = \omega^{d-3}
\ee
where $\omega$ is an integration constant of length dimension equal to one.  We will now consider solutions to the field equation both asymptotically and near the horizon.

First, working asymptotically, we write the metric function as,
\be 
f = f_\infty \frac{r^2}{\ell^2} - \left(\frac{\omega}{r} \right)^{d-3} + \frac{q^2}{r^{2d-6}} + \epsilon h(r) \, .
\ee
We substitute this into the field equations, setting $\epsilon^n = 0$ for $n > 1$ and then setting $\epsilon = 1$ to obtain a linear, inhomogenous, second order differential equation for the correction $h(r)$.  The general structure of this equation is not illuminating, but we note that the homogenous part of the equation, at large $r$, takes the form,
\be 
h'' - \frac{4}{r} h' - \gamma^2 r^{d-3} h = 0
\ee
where
\be 
\gamma^2 = \frac{ 9(d-6)(4d^4 - 49 d^3 + 291d^2 - 514 d + 184)f_\infty^2 \ell^2 \lambda - 4(d-8) \mu f_\infty^3  + 3\ell^6}{ 3 \ell^2  f_\infty (d-1)\left[ 48(d^2 + 5d - 15) \lambda \ell^2 + \mu (d-6) f_\infty  \right] \omega^{d-3}} \, .
\ee
Restricting to the cases where $\gamma^2 > 0$ to avoid solutions which oscillate, we find the approximate solution at large $r$ is given by,
\be 
h \approx A \exp \left(\frac{2 \gamma  r^{(d-1)/2} }{d-1} \right) + B \exp \left( - \frac{2 \gamma r^{(d-1)/2} }{d-1} \right) \, .
\ee
To maintain consistency with the AdS boundary conditions, we must impose that $A = 0$.  We will also see shortly that the second term above is extremely subleading compared to the particular solution.  Thus, we are permitted to drop it as well.  The most useful feature of this analysis is the restriction we have found for the coupling by demanding $\gamma^2 > 0$.  The positivity of the numerator of $\gamma^2$ is equivalent to demanding that the generalized Newton constant is positive (i.e. the graviton is not a ghost), and thus the only additional constraint is,
\be\label{eqn:no_wiggle}
48(d^2 + 5d - 15) \lambda \ell^2 + \mu (d-6) f_\infty  >0 \, .
\ee
In the following analysis, we will demand that this condition hold for all physically reasonable solutions.  It is interesting to note that, given the results presented in~\cite{Hennigar:2017ego} and~\cite{Ahmed:2017jod} for the linearized theories, when either $\lambda = 0$ or $\mu = 0$, the positivity of $\gamma^2$ is equivalent to demanding that the graviton is not a ghost.

Moving on to consider the particular solution, it is a matter of calculation to show that it takes the form,
\begin{align}
h_p(r) = & \frac{P(f_\infty) - 1}{ P(f_\infty)} \left(\frac{\omega}{r} \right)^{d-3} - \frac{P(f_\infty) - 1}{ P(f_\infty)} \frac{q^2}{r^{2d-6}} 
\nn\\
&+ \frac{1}{2} \big[ (d^4 - 8d^3 +13 d^2 - 10d + 32) \mu f_\infty +   2(36d^5 - 147d^4 +1179d^3  
\nn\\
&-5940d^2 +9444d - 3312)\lambda \ell^2 \big]  \frac{P(f_\infty) - 2}{ P(f_\infty)^2} \frac{ f_\infty \omega^{2d-6}}{\ell^4 r^{2d-4}}
\nn\\
&+ \big[(-4d^4 + 42 d^3 - 134 d^2 + 172 d - 104) \mu f_\infty +  (216 d^5 - 342d^4 - 2442d^3 + 5064d^2 
\nn\\
&- 1992d +2016) \lambda \ell^2 \big]   \frac{P(f_\infty) - 2}{ P(f_\infty)^2} \frac{ f_\infty q^2 \omega^{d-3}}{\ell^4 r^{3d-7}}
+ \mathcal{O}\left(\frac{g_1(\lambda, \mu, d)  \, \omega^{3d-9} }{P(f_\infty)^3 r^{3d - 5}}, \frac{g_2(\lambda, \mu, d) \,  q^4}{P(f_\infty)^3 r^{4d - 10}} \right)
\end{align}
where we have given explicit forms for the first four leading corrections and listed the schematic form of the next two corrections to $h_p(r)$. The function $P(f_\infty)$ was defined in eq.~\eqref{eqn:asymp_der} and corresponds to the derivative of the numerical coefficient in the linearized equations of motion. 

Studying the particular solution, we note that the leading order corrections occur at the same orders as the mass and charge, respectively.  This is not unsual for higher curvature gravity, the same occurs in Lovelock gravity if one performs a large $r$ expansion of the metric function.  We also note that the terms appearing in the denominators of the corrections will be positive, non-vanishing quantities.  This is ensured by demanding eq.~\eqref{eqn:no_wiggle} holds, and is actually equivalent to demanding that the graviton is not a ghost in this theory.  Furthermore, we see that, as promised, the particular solution dominates over the homogenous solution, with the latter extremely suppressed.  

Next, we consider the solution near the horizon.  We therefore expand the metric function as,
\be 
f(r) = 4 \pi T (r-r_+) + \sum_{i=2} a_n (r-r_+)^n \, 
\ee
where $T$ is the Hawking temperature of the black brane, which is found easily using the standard Euclidean argument,
\be 
T = \frac{N f'}{4 \pi} \bigg|_{r = r_+} \, .
\ee 
We substitute this ansatz for the metric function into the field equations and demand that it is satisfied order by order in $(r-r_+)$.  This leads to recurrence relationships which determine the series coefficients.  The first two relationships involve only the temperature, mass parameter and horizon radius,
\begin{align}
\label{eqn:nh_cubic}
\omega^{d-3} =& \frac{r_+^{d-1}}{\ell^2} + \frac{q^2}{r_+^{d-3}} - 1024 \pi^3 (d^2 + 5d - 15)\lambda T^3 r_+^{d-4} - 16 \pi^4  (3d-16) \mu  T^4 r_+^{d-5}  \, ,
\nn\\
0 = & \frac{d-1}{\ell^2} r_+^{d-2} - (d-3) \frac{q^2}{r_+^{d-2}} - 4\pi  T r_+^{d-3} + 512 \pi^3 \lambda (d-4)(d^2 + 5d - 15) T^3 r_+^{d-5} 
\nn\\
&+ \frac{16 \pi^4}{3} (d-5)(3d-16) \mu T^4 r_+^{d-6} \, .
\end{align}
These first two terms suffice to give an \textit{exact} description of the thermodynamics.  Going to higher order is straightforward, though the expressions are not illuminating.  At third order in $(r-r_+)$, $a_3$ can be solved for in terms of $a_2$.  This continues to higher order, with all of the series coefficients expressed in terms of the temperature and $a_2$.  This term, $a_2$, is a free parameter whose choice is equivalent to choosing the boundary conditions at infinity.

\begin{figure}[htp]
\centering
\includegraphics[width=\textwidth]{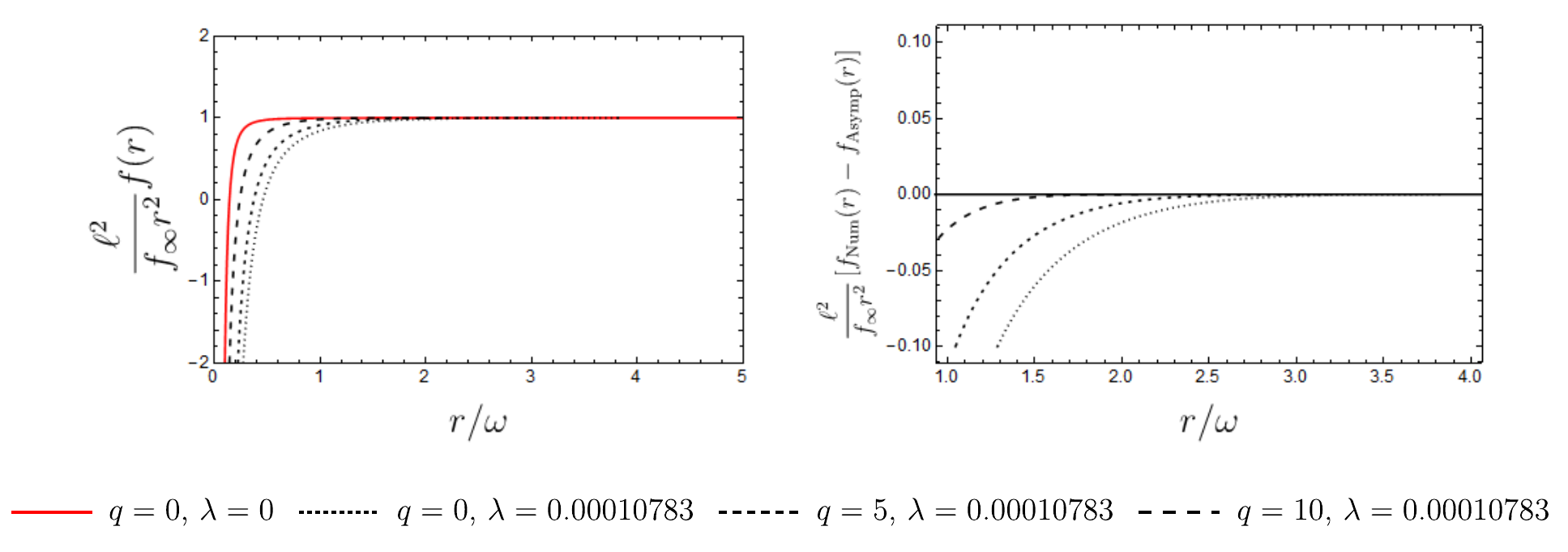}
\includegraphics[width=\textwidth]{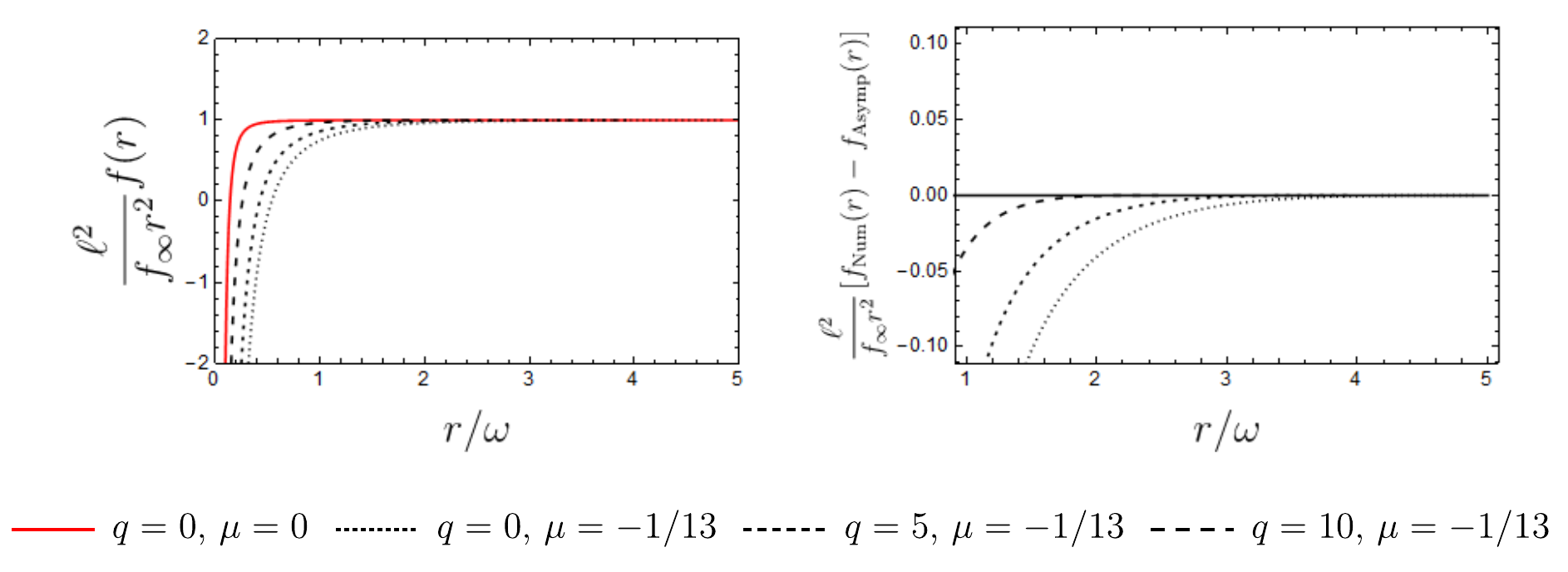}
\caption{{\bf Numerical solutions} (color online). {\it Top row}:  Four dimensional numerical solutions to the cubic theory.  The left panel shows a plot of the rescaled metric function for three different electric charges at non-zero cubic coupling.  The right panel shows the difference between the rescaled metric function and the asymptotic solution as $r$ increases. {\it Bottom row}:  Four dimensional numerical solutions to the quartic theory.  The left panel shows a plot of the rescaled metric function for three different electric charges at non-zero cubic coupling.  The right panel shows the difference between the rescaled metric function and the asymptotic solution as $r$ increases. In all cases, we have set $\ell = 1$.  In each of the left panels, the solid, red line corresponds to the  Einstein-AdS black brane with the choice $\omega = 18.5$.  }
\label{fig:numerical_solutions}
\end{figure}

The near horizon solution can be joined to the asymptotic solution numerically.  The idea is, for some choices of the couplings, electric charge, and horizon radius, to choose a value for $a_2$ and then evaluate the near horizon solution very close to but outside the horizon to obtain values for the metric function and its first derivative.  This is then used as initial data to integrate the field equations numerically.  The constant $a_2 = f''(r_+)$ must be chosen very carefully as to ensure that the proper AdS asymptotics are selected, and we employ the shooting method to do so.  The numerical solution can then be compared to the asymptotic solution, and when the two agree to good precision, the asymptotic solution can be used to continue the solution to infinity.

In general, this is a somewhat challenging to carry out since the differential equation is stiff.  Furthermore, the additional complication in the AdS case is that there are multiple branches for the solution and the correct, ghost-free one must be chosen.  However, for carefully chosen $a_2$ we have found that it is possible to integrate the solution to a radius at which the asymptotic solution [including terms up to $\mathcal{O}(r^{-12})$] is accurate to $1$ in $1,000$ or better.  To make numerical results clearer, we have worked with the rescaled metric function,
\be 
g(r) = \frac{\ell^2}{f_\infty r^2} f(r)
\ee
where $f_\infty$ is chosen to be the root of eq.~\eqref{eqn:asymp_f} which satisfies the condition of no ghosts,  and defined this way $g(r) \to 1$ as $r\to\infty$.  

The results of the numerical investigation are presented in figure~\ref{fig:numerical_solutions} where we have restricted to the case of four dimensions for convenience.  Here we have focused on charged black brane solutions in the cubic and quartic theories for some selections of the electric charge and higher curvature couplings.  The general result in both cases is that the higher curvature couplings push the event horizon outward, while increasing the electric charge pushes it inward.  Along with the numerical solutions, we present plots of the difference between the numerical solution and the asymptotic solution [including terms up to $\mathcal{O}(r^{-12})$].  The numerical solutions were  integrated to large enough radius such that it matches the asymptotic solution up to $1$ part in $1,000$ or better.  From this radius onward, the asymptotic solution can be used to extend the solution to infinity.

We note in closing the interesting  behaviour of the metric function near the origin.  Expanding,
\be 
f(r) = r^s \left(b_0 + r b_1 + r^2 b_2 + \cdots  \right)
\ee
one can show that, in any dimension, a consistent solution is obtained with $s = 0$ and
\be 
f(r) = b_0 + b_2 r^2 + \cdots \, .
\ee
This indicates that, while there is still a curvature singularity at the origin, the singularity is softer than that present in the Einstein gravity case, with the Kretschmann scalar going as $R_{abcd}R^{abcd} \sim r^{-4}$ near the origin (compared to $R_{abcd}R^{abcd} \sim r^{-6}$ when these terms are turned off).  Furthermore, the metric is completely regular at the origin, as opposed to the Einstein gravity case where it is also singular.  It would be interesting to see if the singularity could be completely removed when the theory is coupled to additional matter fields, or to determine if ghosts are required in the linearized spectrum in order to obtain a non-singular black object.

\section{Thermodynamic considerations}
\label{sec:higher_d_thermo}

In this section we will make a number of thermodynamic considerations pertaining to the charged black branes in generalized quasi-topological gravity. We begin by constructing the first law and Smarr relation, followed by discussing various physicality constraints on the higher curvature couplings, and close by presenting the critical behaviour for the black branes.  We shall employ the black hole chemistry formalism~\cite{Kubiznak:2016qmn} here, though we note that the key results can also be observed through different but related methods, e.g. those outlined in~\cite{ChamblinEtal:1999a, Banerjee:2011au, Banerjee:2011raa}.

\subsection{Thermodynamic potentials and first law}

Using the two near horizon relations presented in eq.~\eqref{eqn:nh_cubic}, we can determine the temperature and mass parameter as functions of $r_+$.  Doing so explicitly requires solving a quartic equation, and the final result is not particularly illuminating.  Therefore, we shall work with these equations implicitly.  

The entropy can be calculated using the Iyer-Wald prescription~\cite{Wald:1993nt, Iyer:1994ys} where the entropy is given by,
\be 
S = -2 \pi \oint d^{d-2} x \sqrt{\gamma} \,  E^{abcd} \hat{\varepsilon}_{ab} \hat{\varepsilon}_{cd} 
\ee
where
\be 
E^{abcd} = \frac{\partial \mathcal{L}}{\partial R_{abcd}}
\ee 
and $\hat{\varepsilon}_{ab}$ is the binormal to the horizon, normalized to satisfy $\hat{\varepsilon}_{ab}\hat{\varepsilon}^{ab} = -2$. The integral is evaluated on the horizon of the black hole, which has induced metric $\gamma_{ab}$ and $\gamma = \det \gamma_{ab}$.  In the following we will present the entropy densities, 
\be 
s = \frac{S}{{\rm Vol}\left(\mathbb{R}^{d-2} \right)}
\ee
where ${\rm Vol} \left(\mathbb{R}^{d-2} \right)$ is the (infinite) volume of the submanifold with line element $\sum_{i=1}^{d-2} dx_i^2$.  Carrying out this computation for the action~\eqref{eqn:cubic_action}, we find that the entropy density is given by (see~\cite{Hennigar:2017ego, Ahmed:2017jod} for more details),
\ba
s &=& \frac{r_+^{d-2}}{4}\Bigl[1 -    \frac{384 \pi^2 \lambda}{r_+^2} (d-2)  (d^2 + 5d - 15)  T^2 - \frac{16 \pi^3 \mu }{3 r_+^3} (d-2)(3d-16) T^3 \Bigr]\, .
\ea
Using this expression for the entropy, and defining the pressure in the standard way,
\be 
P = - \frac{\Lambda}{8 \pi} \, ,
\ee
it can be verified that the extended first law holds,
\be 
dM = Tds + V dP + \Phi dQ + \Psi_\lambda d\lambda + \Psi_\mu d\mu
\ee
where the various quantities appearing in this expression are given by,
\begin{align}
M &= \frac{d-2}{16 \pi } \omega^{d-3} 
\, , 
\quad V = \frac{r_+^{d-1}}{(d-1) }
\,, \quad
Q = \frac{\sqrt{2(d-2)(d-3)}}{16 \pi } q
\nn\\
\Phi &= \sqrt{\frac{2(d-2)}{d-3}} \frac{q}{r_+^{d-3}} \, , \quad \Psi_\lambda = 32(d-2)(d^2 + 5d - 15) \pi^2  r_+^{d-4} T^3 \, ,
\nn\\
\Psi_\mu &=  \frac{\pi^3}{3} (d-2)(3d-16)  r_+^{d-5} T^4
\end{align}
and we note that the above quantities (except $\Phi$) are densities, that is, defined up to factors of the (infinite) volume of the transverse space.  The above quantities satisfy the Smarr relation that follows from Eulerian scaling,
\be 
(d-3)M = (d-2) Ts - 2 PV + (d-3) \Phi Q + 4 \lambda \Psi_\lambda + 6\mu \Psi_\mu \,.
\ee
We also note that we recover the expected relationship for a $(d-1)$-dimensional CFT with chemical potential, 
\be 
M = \frac{d-2}{d-1} \big[Ts + \Phi Q \big] \, .
\ee

To construct the equation of state, one can simply rearrange the second equation in eq.~\eqref{eqn:nh_cubic} for pressure, obtaining,
\be 
P = \frac{T}{v}  -   (d-4) \beta_3 \frac{ T^3 }{v^3} - (d-5) \beta_4 \frac{T^4}{v^4} + \frac{e^2}{v^{2d-4}} \,.
\ee
Here, for the sake of simplicity, we have defined,
\begin{align}\label{eqn:rescaled_thermo} 
v &= \frac{4 r_+}{ (d-2)} \, ,\quad \lambda = \frac{(d-2)^2}{2048 \pi^2 (d^2 + 5d - 15)} \beta_3 \, , \quad \mu = \frac{3 ( d-2)^3}{256 \pi^3 (3d - 16)} \beta_4 \, , 
\nn\\
e^2 &= \frac{16^d (d-3) q^2 }{4096 \pi (d-2)^{2d-5}}
\end{align}
as the specific volume, rescaled couplings, and rescaled charge, respectively.  The equation of state here is non-linear (as opposed to linear) in the temperature.  This non-linear dependence on the temperature was first observed for Einsteinian cubic gravity in~\cite{Hennigar:2016gkm}, and seems to be a general feature for black objects in the generalized quasi-topological class of theories.\footnote{As a result of the non-linearity in temperature, the general results presented in~\cite{Majhi:2016txt} do not directly apply, though we expect that a simple modification of that argument could extend it to this more general case, though one might expect additional possibilities for the critical exponent $\alpha$.} We note that, in four dimensions, the cubic generalized quasi-topological term does not contribute to the equation of state and similarly for the quartic term in $d = 5$.   We will consider in a later section the effect of including even higher order generalized quasi-topological terms in four dimensions.

In closing, we note that the Gibbs free energy density is given in the standard way by, $G = M - Ts$
which has the explicit form,
\be 
\mathcal{G} = \left[\frac{4}{d-1} \right]^{d-1} G = \frac{v^{d-1}P}{d-1} - \frac{v^{d-2} T}{d-2} + \frac{e^2}{(d-3) v^{d - 3}} + \beta_3 v^{d-4} T^3 + \beta_4 v^{d-5} T^4 
\ee
The state of the system is taken to be that which minimizes the Gibbs free energy at constant temperature and pressure.  In defining $\mathcal{G}$ we have absorbed some inconvenient (but positive) factors to make the expression easier to work with.  

\subsection{Physicality conditions}

Before discussing the critical behaviour of the black branes, we discuss a number of conditions which the coupling constant must meet in order for the results to be physically reasonable.  

The first condition we impose is that the effective Newton's constant has the correct sign, which is equivalent to demanding that the graviton is not a ghost.  This condition can be read off from the pre-factor appearing in the linearized equations of motion about the AdS background, and we refer the reader to~\cite{Hennigar:2017ego} for the details this calculation.  The result is that the coupling must satisfy the following inequality,
\be 
P(f_\infty) =  1 + 3 \frac{\lambda}{\ell^4} (d-6)(4d^4 - 49 d^3 + 291 d^2 - 514 d + 184) f_\infty^2 - \frac{4 \mu }{3 \ell^6} (d-8) f_\infty^3 > 0 \, .
\ee
In this inequality, the value of $f_\infty$ is obtained by solving eq.~\eqref{eqn:asymp_f}, and it must be positive for AdS asymptotics.
It is noteworthy that, in the cases where either $\lambda = 0$ or $\mu =0$,  this is the same constraint that eliminates the oscillating asymptotic solutions.  Recasting the ghost constraint in terms of the rescaled thermodynamic variables presented in eq.~\eqref{eqn:rescaled_thermo} we obtain,
\be 
1+\frac{3}{8} {\frac { \left( 4{d}^{4}-49{d}^{3}+291{d}^{2}-514d+184 \right)  \left( d-6 \right) f_\infty^{2} \beta_3 {P}^{2}}{\left( d-1 \right) ^{2}  \left( {d}^{2}+5d-15 \right)  }} - \frac{64(d-8) f_\infty^3 \beta_4  P^3}{(d-1)^3(3d-16)} > 0 \, .
\ee
In general, this inequality is satisfied only by the Einstein branch of the theory.  That is, the branch which has a smooth $\lambda \to 0$ limit.  Additionally, to maintain consistency with AdS asymptotics and avoid oscillating parts of the asymptotic solution, we enforce eq.~\eqref{eqn:no_wiggle} which reads,
\be 
\beta_3 + \frac{8(d-6) f_\infty P}{(d-1)(3d-16)} \beta_4 > 0\, .
\ee

The third condition we impose is that the entropy is positive.  It is a general feature of higher curvature theories that the black hole entropy can be negative for certain values of the coupling constants.  While there are often no other pathologies associated with the negative entropy solutions, a number of authors have speculated that these solutions may be unstable~\cite{Cvetic:2001bk, Nojiri:2001pm}.  Here, we will impose that the entropy should be positive, but remark what happens if this condition were relaxed.   In terms of the rescaled thermodynamic parameters, the entropy will be positive provided that,
\be 
s > 0 \Rightarrow 1 - 3 (d - 2)\beta_3 \frac{T^2}{v^2} - 4 (d - 2)\beta_4  \frac{T^3}{v^3} > 0 \, .
\ee 

Since the temperature and specific volume will always be positive for sensible solutions, this inequality is satisfied trivially if the coupling is negative.  When the coupling is positive, enforcing this is more cumbersome, since the temperature must  be solved for in terms of the pressure and specific volume through the equation of state.

\subsection{Critical behaviour}

We now turn to a discussion of the critical behaviour.  The equation of state,
\be 
P = \frac{T}{v}  -   (d-4) \beta_3 \frac{ T^3 }{v^3} - (d-5) \beta_4 \frac{T^4}{v^4} + \frac{e^2}{v^{2d-4}} \,,
\ee
from which it is clear that the term arising from the electric charge dominates for small $v$ (i.e. small $r_+$), in all dimensions larger than four, but is comparable to the quartic term in four dimensions. The equation of state will have a critical point provided that
\be 
\frac{\partial P}{\partial v} = \frac{\partial^2 P}{\partial v^2} = 0 \, .
\ee
We find that, including terms up to quartic order, there are no points in the parameter space which satisfy this condition in four dimensions (we shall return to the four dimensional case in a later section).

Next we consider the five dimensional case. Here, calculations are simplified due to the fact that in five dimensions the quartic term drops out of the equation of state.  Therefore, we will consider this case in detail as an illustrative example.  

In five dimensions, there is a critical point given by the following critical values,
\be 
v_c = \frac{\sqrt{3} \beta_3^{1/12} e^{1/3} }{5^{1/12}} \, , \quad T_c = \frac{5^{5/12} e^{1/3}}{\sqrt{3} \beta_3 ^{5/12}} \, , \quad P_c = \frac{5 \sqrt{5}}{27 \sqrt{\beta_3}} \, .
\ee
The ratio of these values is independent of the properties of the black brane and is therefore  a universal property of the theory,
\be 
\frac{P_c v_c}{T_c} = \frac{5}{9} \, .
\ee
It is interesting to note that this ratio is distinct from the value obtained for spherical, charged black holes in five dimensional Einstein gravity~\cite{Gunasekaran:2012dq}, while the ratio for four dimensional spherical black holes in Einsteinian cubic gravity was found to coincide with the corresponding value in Einstein gravity~\cite{Hennigar:2016gkm}.  These differences suggest a connection between the ratio of critical quantities and the topology of the horizon cross-sections.  

In five dimensions, the regions where each physicality constraint holds can be solved for analytically at the critical point.  The resulting expressions are rather complicated and not particularly illuminating in some cases, thus we will display approximate numeric values in those cases.  The graviton will be a ghost at the critical point, if
\be 
 \beta_4 >  1.309 \beta_3^{3/2} \, .
\ee 
The solution becomes inconsistent with the asymptotic AdS boundary conditions and oscillates if,
\be 
\beta_4 < -1.226 \beta_3^{3/2} \, .
\ee
The solution satisfies all physicality constraints provided that the coupling is in the range,
\be 
-1.226 \beta_3^{3/2} < \beta_4 < -\frac{9 \sqrt{5}}{25} \beta_3^{3/2} \, ,
\ee
and elsewhere the entropy is negative, but the solution is otherwise well-defined.  It is important to note that the entropy will always be negative at the critical point if the quartic coupling is set to zero.  

\begin{figure}[htp]
\centering
\includegraphics{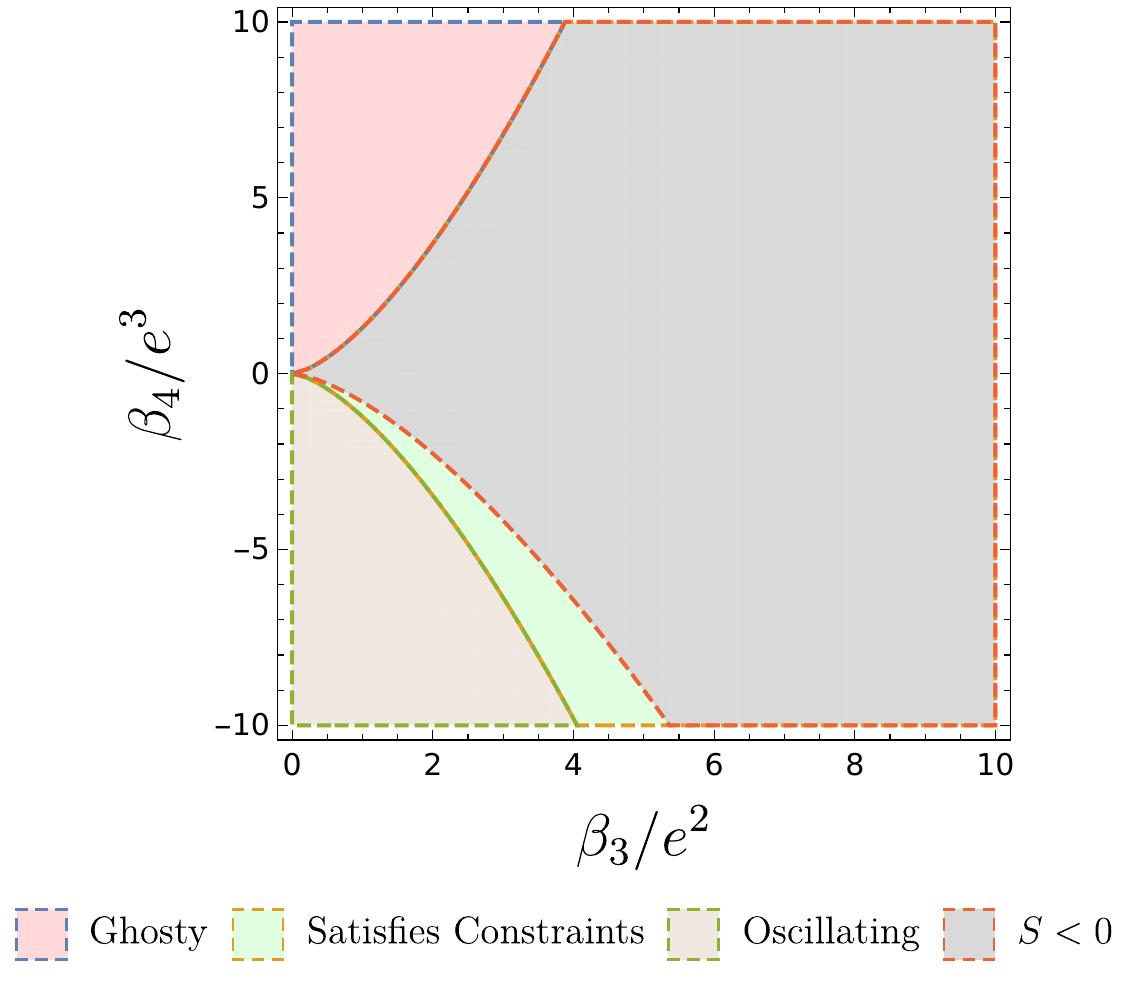}
\caption{{\bf Constraint profiles in five dimensions}  (color online).  This plot shows a representative sampling of the parameter space, highlighting where each physicality constraint holds.  The dimensionful quantities have been rescaled in terms of the electric charge parameter, $e$, which was then set to unity. }
\label{fig:5d_constraints}
\end{figure}

The conclusion of this is that there is a large portion of the parameter space for five dimensional black branes where all physicality constraints hold and phase transitions occur.  This space becomes even larger in the case where negative entropy solutions are not discarded.  The various physicality conditions are illustrated graphically in figure~\ref{fig:5d_constraints}.

The critical points are characterized by mean field theory critical exponents,
\be 
\alpha = 0 \, , \quad  \beta = \frac{1}{2} \, , \quad \gamma = 1 \, , \quad \delta = 3 \,,
\ee
as can be easily confirmed~\cite{Gunasekaran:2012dq} by studying the equation of state near the critical point\footnote{Dropping terms such as $\tau^2$ from the near critical expansion while keeping terms like $\phi^3$ is justified since these terms drop out in the calculations of the critical exponents.  See~\cite{Gunasekaran:2012dq} for similar discussion.},
\be 
\frac{P}{P_c} = 1 - \frac{6}{5} \tau  + \frac{36}{5} \phi \tau - 3 \omega^3 + \mathcal{O}(\tau^2, \phi^2\tau,  \phi\tau^2)
\ee
where
\be 
v = v_c(\phi + 1) \, , \quad T = T_c (\tau + 1) \, .
\ee

\begin{figure}[htp]
\centering
\includegraphics[width=0.7\textwidth]{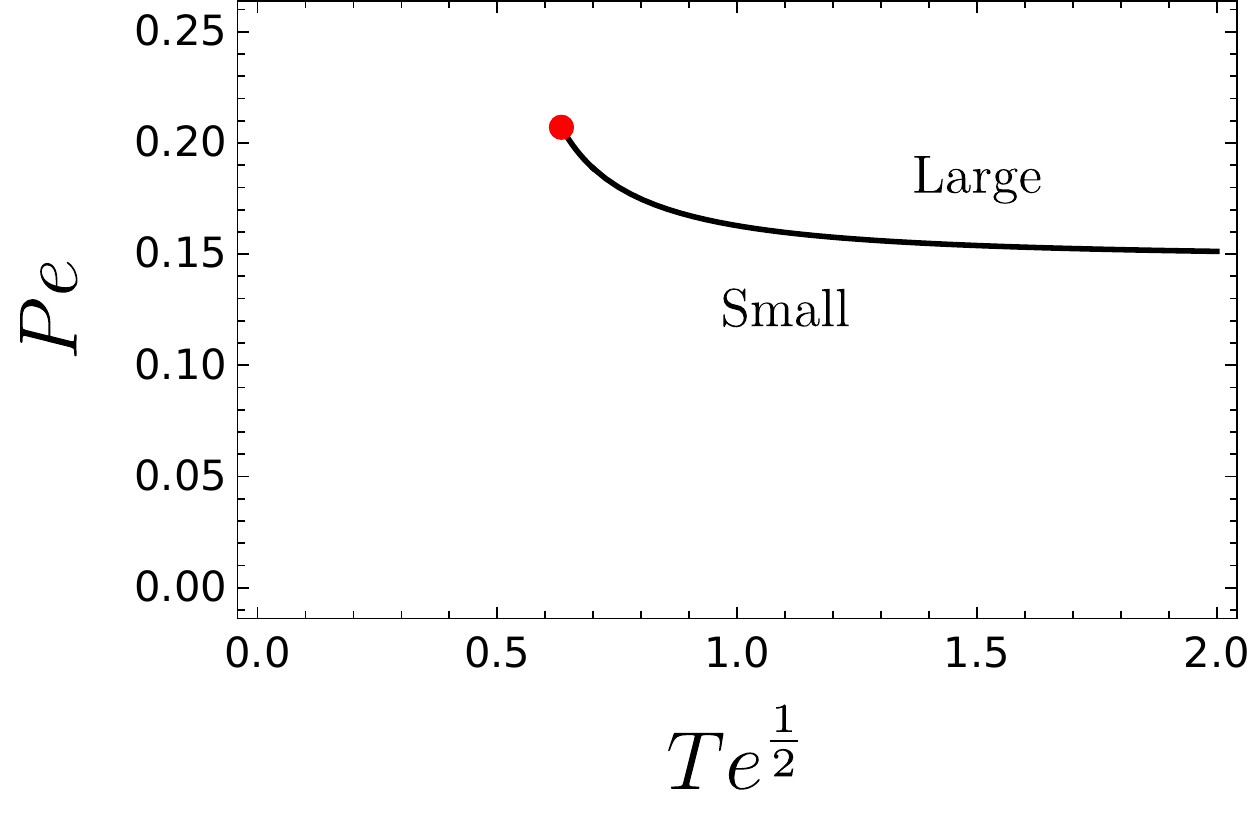}
\caption{{\bf Phase diagram } (color online).  The phase diagrams for five dimensional charged  black branes is displayed.  This plot was constructed for the particular example of $\beta_3 = 4$ and $\beta_4 = -9$, though these values are not special and any combination in the green region of figure~\ref{fig:5d_constraints} would produce a qualitatively similar plot.  Here the dimensionful quantities have been rescaled in terms of the electric charge parameter $e$, which has subsequently been set to unity.}
\label{fig:5d_phase_diagram}
\end{figure}

The critical point marks the starting point of a first order phase transition, with distinct phases (large and small) appearing at higher temperatures.  This is in contrast to typical van der Waals behaviour where the critical point marks the end point of a first order phase transition, with distinct phases appearing at lower temperatures, as shown in figure~\ref{fig:5d_phase_diagram}.  At high tempertures, the coexistence curve flattens, lying along a constant pressure.  This indicates that, while at low temperatures, an increase in temperature can lead to a phase transition at constant pressure, at higher temperatures the black brane is either small or large at constant pressure, and a phase transition can only be brought about by changing the pressure.

\begin{figure}[htp]
\centering
\includegraphics[width=0.45\textwidth]{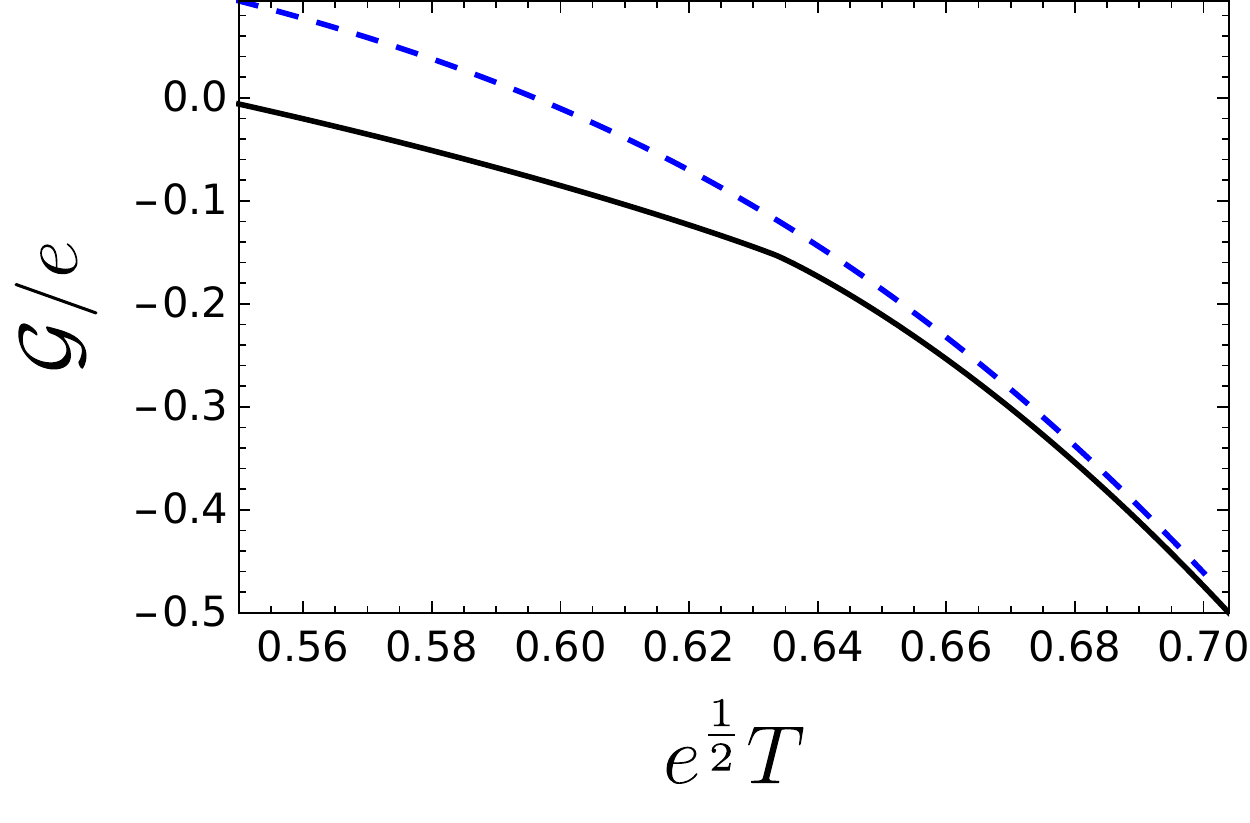}
\includegraphics[width=0.45\textwidth]{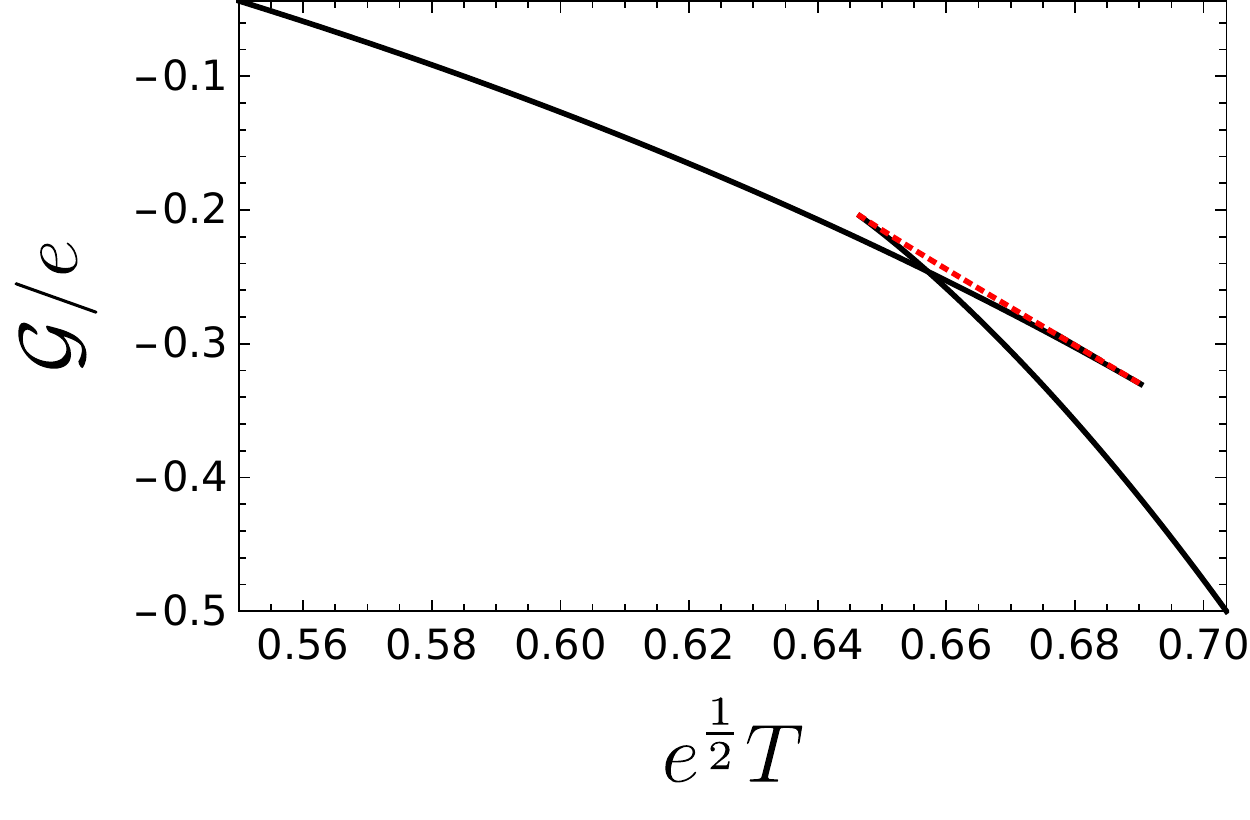}
\includegraphics[width=0.45\textwidth]{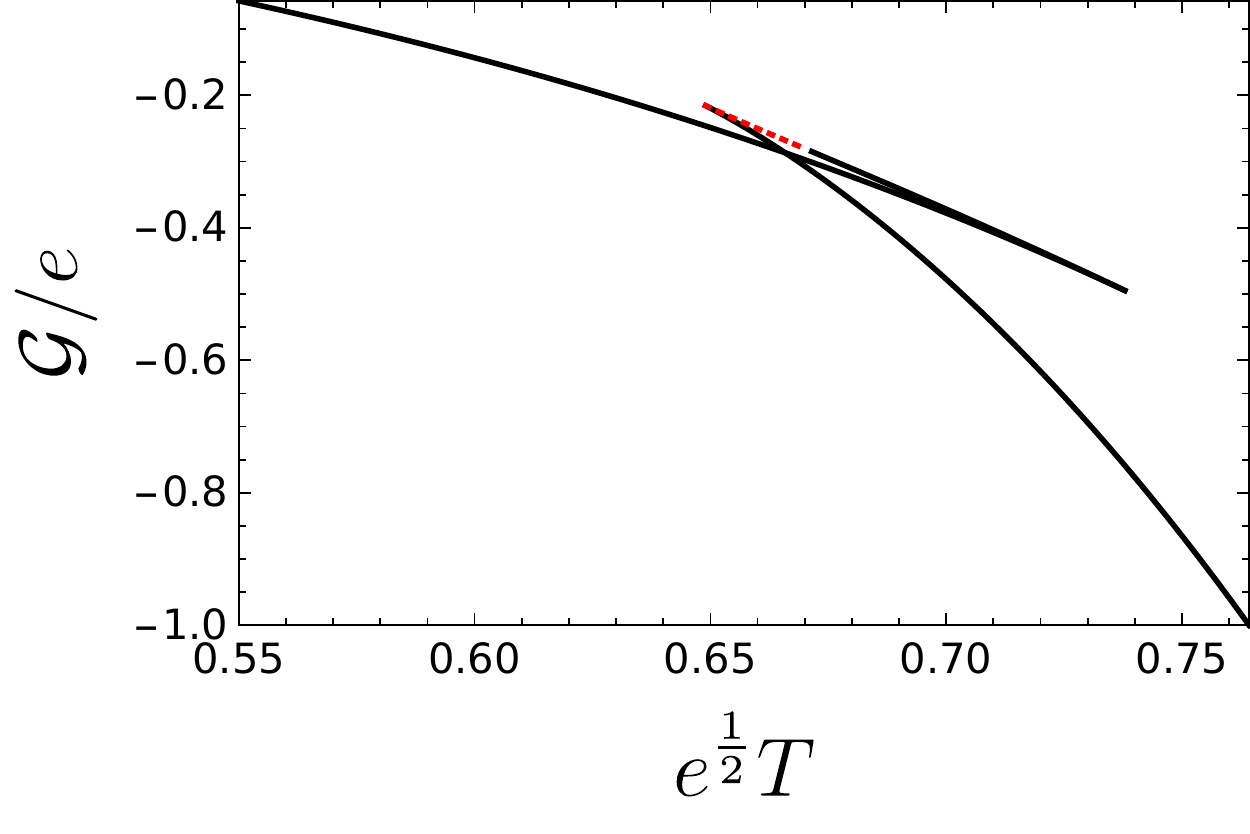}
\includegraphics[width=0.45\textwidth]{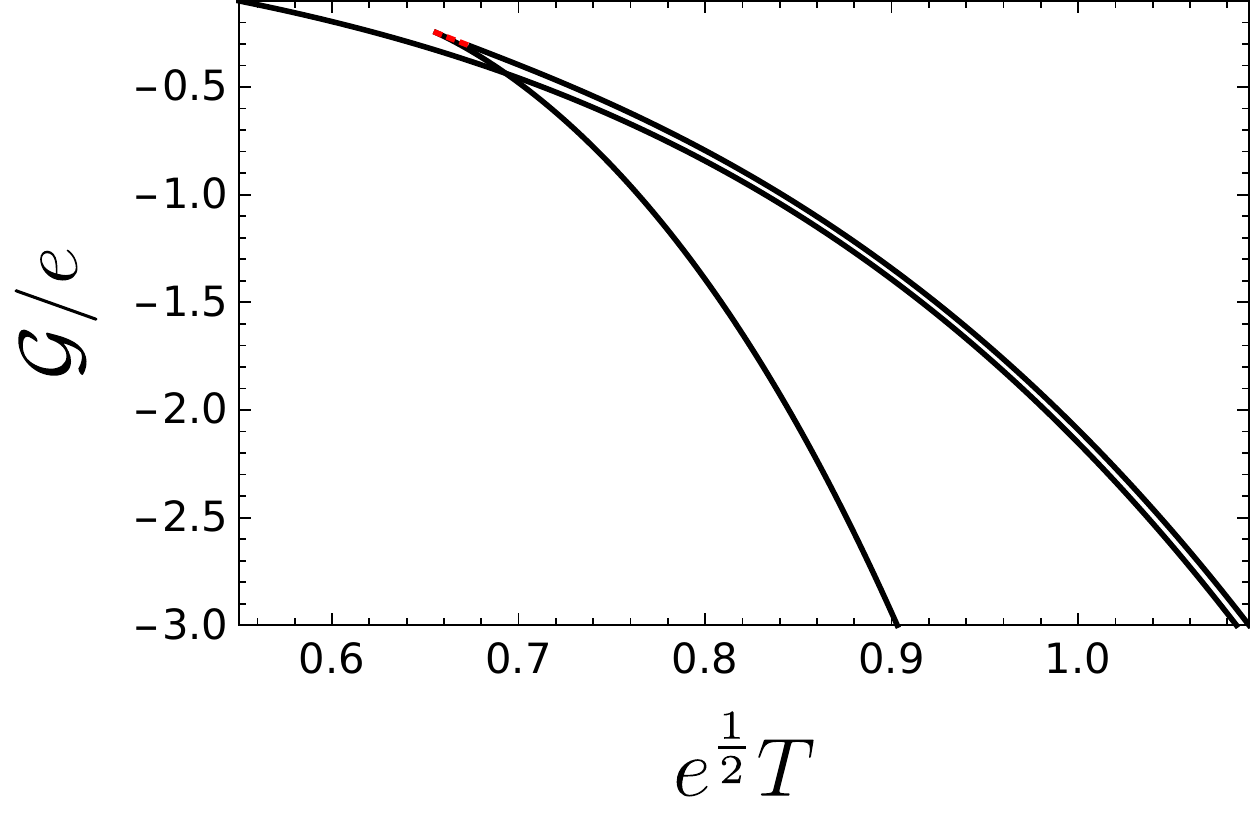}
\caption{{\bf Free energy in five dimensions} (color online).  {\it Top left}: Plot of the Gibbs free energy for $P = P_c$ (solid, black curve) and for $P \approx 1.21 P_c$ (dashed, blue curve).  {\it Top right}: Plot of the Gibbs free energy for $P \approx 0.96 P_c$. {\it Bottom left}: Plot of the Gibbs free energy for $P \approx 0.95 P_c$. {\it Bottom right}: Plot of the Gibbs free energy for $P \approx 0.92 P_c$.  In each plot, the dotted, red lines indicate portions of the curves where the specific heat is negative.  This plot has been constructed for $\beta_3 = 4$ and $\beta_4 = -9$, with all physicality conditions enforced.  The dimensionful quantities have been rescaled in dimensions of the electric charge parameter $e$, which has then been set to unity in these plots. }
\label{fig:5d_free_energy}
\end{figure}

Near the critical point, the structure of the free energy is particularly rich and displays some novel features, as illustrated in figure~\ref{fig:5d_free_energy}. When the pressure exceeds the critical pressure, the free energy has a single branch and their are no phase transitions.  At the critical pressure, the free energy exhibits a kink corresponding to the critical point.  These are standard features associated with all examples of typical van der Waals behaviour.  However, when the pressure dips below its critical value, interesting features emerge.  For pressures close to but below the critical pressure, the Gibbs free energy displays the standard swallowtail typical for van der Waals behaviour.  The black branes which correspond to the parameter values on the swallowtail portion of the Gibbs free energy have negative specific heat, as can be inferred from the curvature of the Gibbs free energy and recalling that,
\be 
C_p = - T \frac{\partial ^2 G}{\partial T^2} \, .
\ee 
The negative specific heat is indicated graphically in figure~\ref{fig:5d_free_energy} by red, dotted lines.  As the pressure decreases further, the swallowtail rapidly grows, until eventually it no longer closes at finite temperature.  At the same time, a stable portion ($C_P > 0$) emerges on the swallowtail, which grows in size until, at low enough pressure, effectively the entire upper branch of the swallowtail corresponds to positive specific heat.  Graphically, this can be seen by the red, dotted portions of the curve shrinking smaller and smaller from the top right to bottom right plots of figure~\ref{fig:5d_free_energy}.  Physically, this means that there are, for large portions of the parameter space, three thermodynamically stable black branes, though only one of these minimizes the free energy.  It would be interesting to see if this results extends to spherical or topological black holes in the theory.

To close this section, we will make some remarks about criticality in higher dimensions.  In dimensions larger than five, it is both the cubic and quartic terms that will contribute to the equation of state, and a critical point governed by mean field theory exponents is a generic feature.  To study the critical point then requires one to solve a quartic equation, which is of course possible, but the result is not illuminating.  Setting either $\beta_3$ or $\beta_4 = 0$ results in a system with a far simpler solution.  In these cases, there exists a single critical point that can be studied analytically yielding critical ratios,
\begin{align}
\frac{P_c v_c}{T_c} &= \frac{1}{3} \frac{2d-5}{d-2} \quad \text{ if $\beta_4 = 0$},
\nn\\
\frac{P_c v_c}{T_c} &= \frac{3}{8} \frac{(d-2)(d-5)^{1/3}(2d-9)(d-4)^{1/3}(d^2 - 9d + 20) + 2(d^2 - 9d + 20)}{(d-4)^{4/3}(d-5)^{1/3} (d - 2) (d^2- 9d + 20)^{2/3}} \quad \text{ if $\beta_3 = 0$},
\end{align} 
but then one is left with the problem that the entropy is always negative at the critical point:
\begin{align}
s_c &\propto - \frac{6(d-3)}{(2d-7)(d-4)} \quad \text{ if $\beta_4 = 0$},
\nn\\
s_c &\propto - \frac{3}{2} \frac{3d-10}{(d-4)(d-5)} \quad \text{ if $\beta_3 = 0$}.
\end{align}

\begin{figure}[htp]
\centering
\includegraphics[width=\textwidth]{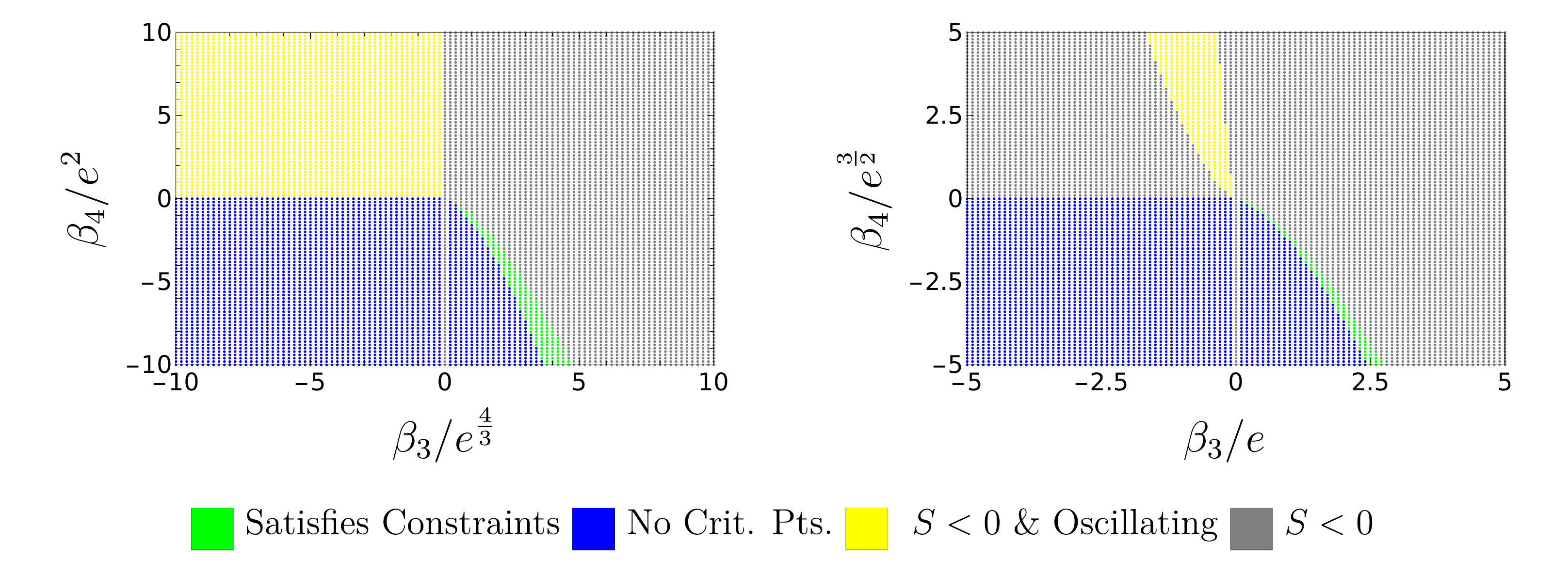}
\\ 
\vspace{15 pt}
\includegraphics[width=0.48\textwidth]{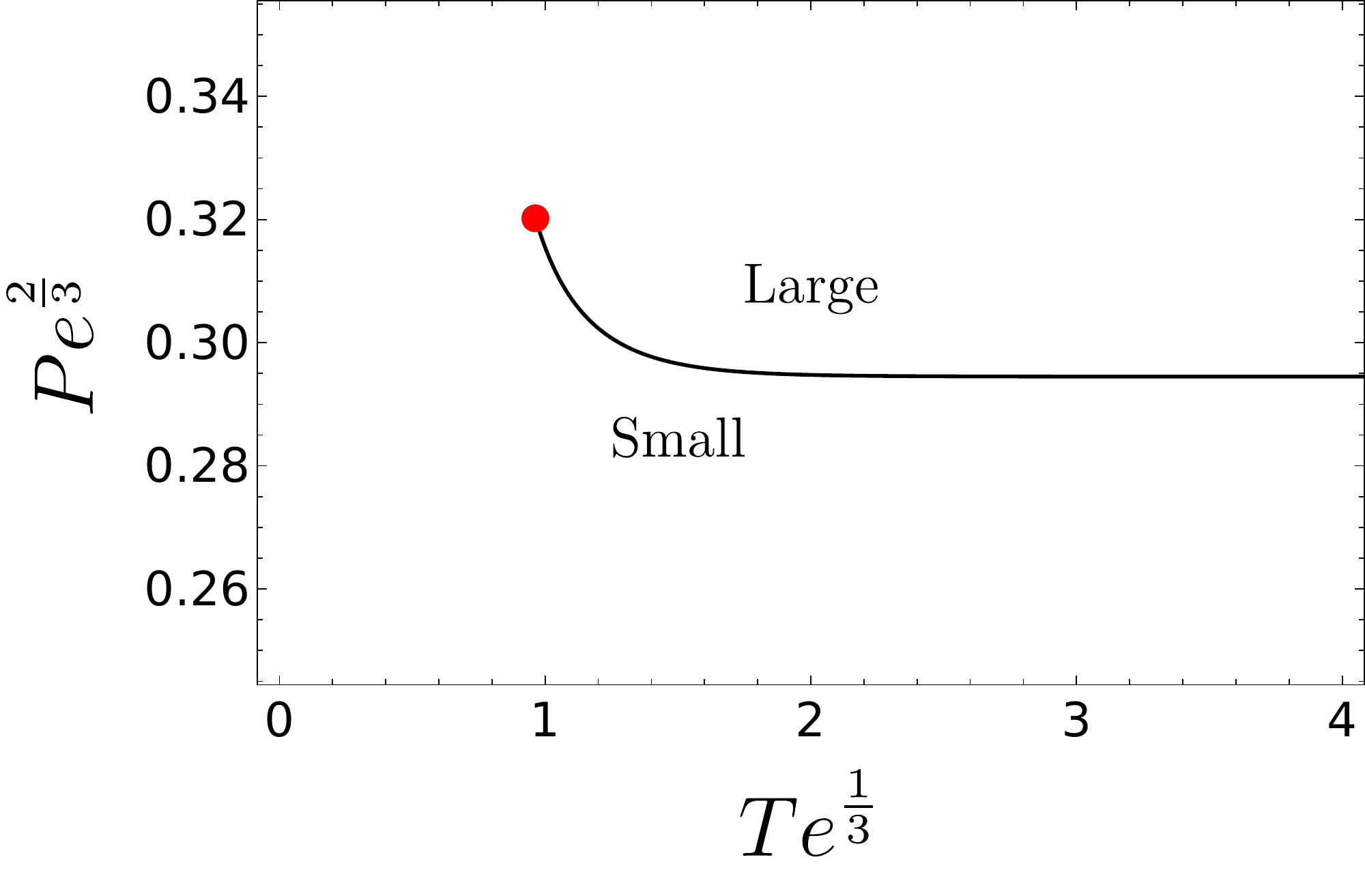}
\quad
\includegraphics[width=0.48\textwidth]{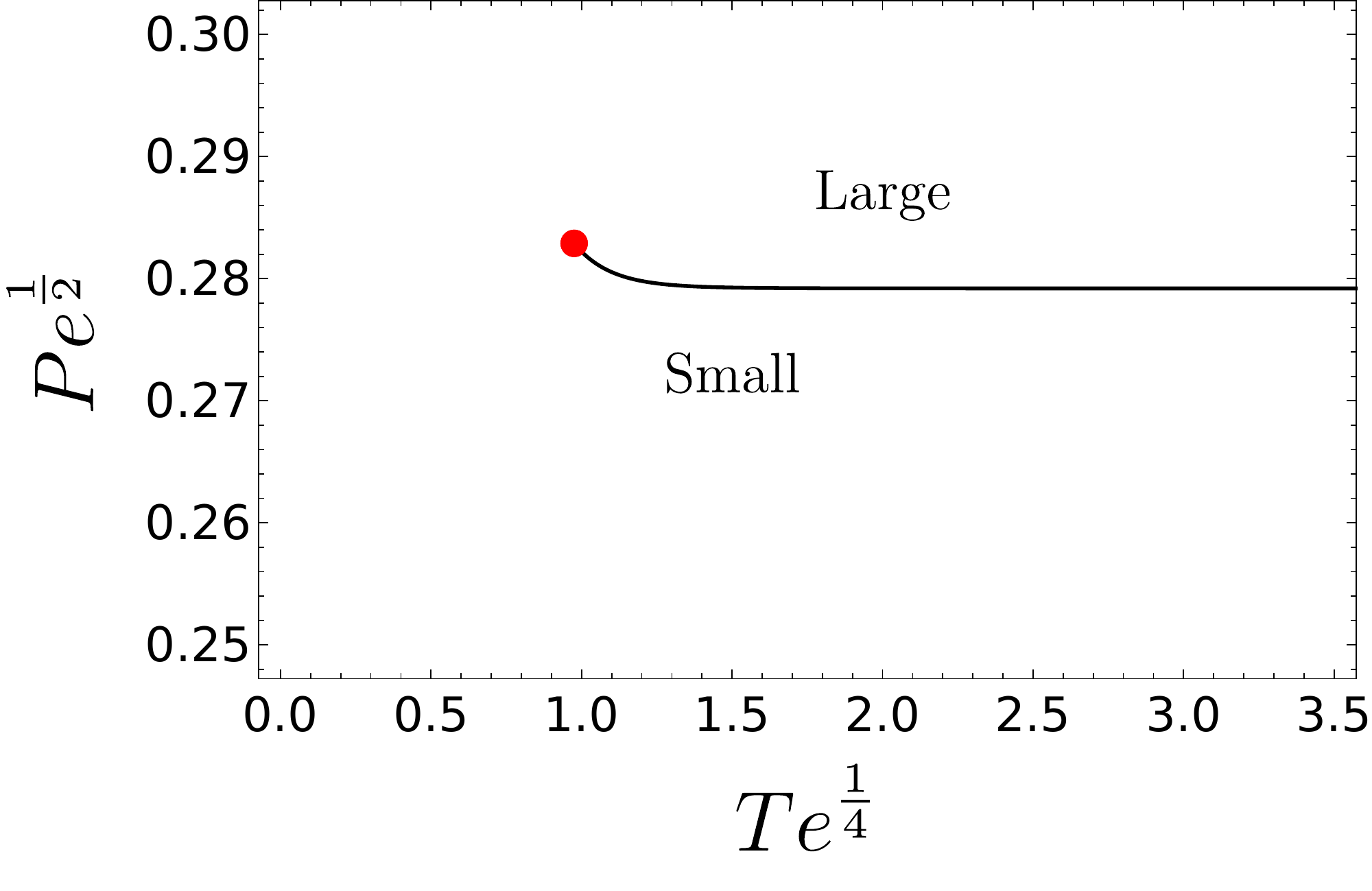}
\caption{{\bf Physicality constraints and phase diagrams in higher dimensions} (color online). {\it Top left}: The  constraints  for six dimensions. {\it Top right}: The constraints for seven dimensions.  {\it Bottom left}: The phase diagram for six dimensions.  The red dot represents the critical point, while the black line represents the coexistence curve for a first order phase transition.  This plot has been constructed for the particular choice of $\beta_3 = 1, \beta_4 = -1$ in units of the electric charge parameter, $e$. {\it Bottom right}: The phase diagram for seven dimensions.  The red dot represents the critical point, while the black line represents the coexistence curve for a first order phase transition.  This plot has been constructed for the particular choice of $\beta_3 = 1, \beta_4 = -5/4$ in units of the electric charge parameter, $e$. }
\label{fig:constraints_higher_d}
\end{figure}

More physically interesting scenarios arise if both the cubic and quartic contributions are non-trivial. To illustrate the behaviour in this case, we will proceed by presenting a few illustrative numerical studies in particular dimensions.  The first consideration to be made involves the study of the physicality constraints in higher dimensions.  Investigating the parameter space numerically in various dimensions reveals a structure similar to that shown in figure~\ref{fig:5d_constraints}.  We show the results explicitly for six and seven dimensions in figure~\ref{fig:constraints_higher_d}.  The notable feature is that, as the spacetime dimension increases, the portion of parameter space in which the critical point is physical gradually shrinks, but remains of non-zero size.  

The phase diagrams resulting from the critical behaviour in higher dimensions are qualitatively similar to the five dimensional case, and we show the results for six and seven dimensions in figure~\ref{fig:constraints_higher_d} for the sake of completeness.  In each case, the Gibbs free energy exhibits behaviour qualitatively similar to that seen in five dimensions and illustrated in figure~\ref{fig:5d_free_energy}. 

\section{Phase transitions in four dimensions?}
\label{sec:four_d}
In the previous section, we have seen that both the cubic and quartic generalized quasi-topological terms induce phase transitions for black branes in all dimensions larger than four. However, in four dimensions neither of these terms --- nor combinations of them --- lead to phase transitions.  It is natural to wonder, then, if the four dimensional branes are immune to criticality or if the inclusion of even higher order terms can induce criticality.

Luckily, the structure of the field equations of the GQT terms have been worked out recently in four dimensions to arbitrary order in the curvature, and explicit examples of the Lagrangians were provided up to  $10^{th}$ order in the curvature~\cite{Bueno:2017qce}.  Here we shall import the results of that study and investigate the consequences for four dimensional black branes. 

In four dimensions, the action for the theory containing representatives from each order in the curvature is given by,
\begin{equation}
\mathcal{I}= \frac{1}{16\pi G} \int d^4 x \sqrt{-g} \left[ \frac{6}{\ell^2} + R +\sum_{n=3}^{\infty}\gamma_n  \mathcal{S}_{(n), 4} - \frac{1}{4} F_{ab}F^{ab} \right]\, ,
\end{equation}
where the terms $\mathcal{S}_{(n), 4}$ have been determined up to tenth order in the curvature ($n=10$) in~\cite{Bueno:2017qce} --- cf. appendix A of that work.  The field equations which follow from this action can be worked out at any order in the curvature by inferring the dependence from the first few of these terms.  The result for the field equations in the case $k=0$ is~\cite{Bueno:2017qce},
\begin{align}
&\frac{r^3}{\ell^2} -r f-\sum_{n=3}^{\infty}\gamma_n\left(\frac{f'}{r}\right)^{n-3}\Bigg[\frac{f'^3}{n}+\frac{(n-3)f}{(n-1)r}f'^2-\frac{2}{r^2}f^2 f'-\frac{1}{r}f f''\left(r f'-2f\right)\Bigg] = \omega - \frac{q^2}{r} \, 
\label{kfequationn2} 
\end{align}
where we have included a Maxwell field with the same conventions used as in section~\ref{sec:charged_BB}.  Note that for the cubic and quartic terms, the convention for the higher curvature couplings here are related to those used earlier by,
\begin{align}
\gamma_3 &= 1008 \lambda \, , \quad  \gamma_4 = - \mu \, .
\end{align}

Considering black brane solutions, far from the horizon the metric takes the form,
\be 
f(r) = f_\infty \frac{r^2}{\ell^2} - \frac{\omega}{r} + \frac{q^2}{r^2}  + h(r)
\ee
where $f_\infty$ solves the polynomial equation,
\be 
1 - f_\infty + \sum_{n=3}^{\infty} \gamma_n  \frac{2^{n} }{n(n-1)} \frac{f_\infty^n}{\ell^{2n - 2}} = 0 \, .
\ee
The leading order corrections can be obtained in the same way as in section~\ref{sec:charged_BB}, and we find at large $r$, that $h(r)$ satisfies an inhomogenous second order differential equation.  The homogenous part of this reads,
\begin{align}
h'' - \frac{4}{r} h' - \frac{1}{3} \frac{P(f_\infty)}{\sum_{n=3}^\infty \gamma_n (2 f_\infty)^{n-3} \ell^{4 - 2n}} \frac{r}{f_\infty \omega} h = 0 \, . 
\end{align}
where $P(f_\infty)$ is given by,
\be 
P(f_\infty) = 1 -  \sum_{n=3}^{\infty} \gamma_n  \frac{2^{n} }{n(n-1)} \frac{f_\infty^{n-1}}{\ell^{2n - 2}} \, 
\ee
and $P(f_\infty) > 0$ ensures that the asymptotic AdS region is free from ghosts, i.e. ghosts are not present in the dual CFT.  Additionally, we see that the solution will not oscillate provided that,
\be 
\sum_{n=3}^\infty \gamma_n (2 f_\infty)^{n-3} \ell^{4 - 2n} > 0 \, .
\ee

Solving the homogeneous equation as before, and also determining the particular solution, we have
\begin{align}
h(r) &\approx A \exp \left(- 2 \sqrt{\frac{1}{3 f_\infty \omega} \frac{P(f_\infty)}{\sum_{n=3}^\infty \gamma_n 2^{n-3} \ell^{4 - 2n}}}  r^{3/2} \right) -  \frac{1- P(f_\infty)}{P(f_\infty)} \frac{\omega}{r} + \frac{1- P(f_\infty)}{P(f_\infty)} \frac{q^2}{r}
\nn\\
&+ \frac{1}{ P(f_\infty)^2} \left( f_\infty + \sum_{n=3}^\infty \gamma_n \frac{2^n f_\infty^{n}}{(n-1) \ell^{2n-2}} \right) \left(\sum_{n=3}^\infty \frac{\gamma_n}{\ell^{2n-2}} f_\infty^{n-3}  \right) \left[ - \frac{7}{4} \frac{  \omega^2 \ell^2 }{r^4}  + \frac{13}{2} \frac{ \omega q^2 \ell^2}{r^5} \right] 
\nn\\
&+ \mathcal{O}\left(\frac{1}{r^6} \right)
\nn\\
\end{align}
where, as before, we have set the coefficient of the growing exponential mode to zero for consistency with the AdS boundary conditions.  As expected, the exponential term is extremely subleading, and can be neglected in the asymptotic expansion.  The black branes are then characterized completely by the single integration constant, $\omega$ --- there is no higher derivative hair.

Using the same ansatz as in section~\ref{sec:charged_BB}, the field equations can be solved near the horizon.  The first two relationships, which suffice to completely and exactly characterize the thermodynamics, are found to be,
\begin{align}
\omega &= \frac{r_+^3}{\ell^2} + \frac{q^2}{r_+} - \sum_{n=3}^{\infty} \gamma_n\frac{r_+^3}{n} \left(\frac{4 \pi T }{r_+} \right)^n \, ,
\nn\\
0 &= 3 \frac{r_+^2}{\ell^2} - \frac{q^2}{r_+^2} - 4 \pi r_+ T   - \sum_{n=3}^\infty  \gamma_n  \left[ \frac{(n-3)r_+^2}{n(n-1)} \right] \left(\frac{4 \pi T }{r_+} \right)^n \, .
\end{align}
The higher order terms are easily obtained, but are not illuminating.  As usual, all higher order terms can be expressed in terms of the single free parameter, $a_2$ in the near horizon expansion.  The choice of this parameter is equivalent to the choice of the boundary conditions at infinity.

The four dimensional black branes satisfy the first law of thermodynamics,
\be 
dM = Tds + VdP + \Phi dQ + \sum_{n=3}^{\infty} \Psi_n d \gamma_n
\ee
with the following thermodynamic quantities,
\begin{align}
s &= \frac{r_+^2}{4}\left[1 - 2\sum_{n=3}^\infty \frac{\gamma_n}{n-1} \left(\frac{4 \pi T }{r_+} \right)^{n-1}  \right] \, , \quad \Phi =  \frac{2 q}{r_+} \, , \quad Q = \frac{q}{8 \pi  } \, , 
\nn\\
M &=  \frac{\omega}{8 \pi  } \, , \quad V = \frac{r_+^3}{3 } \, , \quad \Psi_n = \frac{r_+^3}{8 \pi  n(n-1)} \left(\frac{4 \pi T }{r_+} \right)^{n} \, .
\end{align}
Note that the above quantities, with the exception of the electric potential, are defined as densities, removing the (infinite) transverse volume, ${\rm Vol}\left(\mathbb{R}^{2} \right)$. The thermodynamic quantities obey the Smarr formula which is derived from Eulerian scaling,
\be 
 M = 2 Ts - 2 PV +  \Phi Q + \sum_{n=3}^{\infty} 2(n - 1) \Psi_n \gamma_n \, 
\ee
and also obey
\be 
M = \frac{d-2}{d-1} \left[ Ts + \Phi Q \right]
\ee
as expected for a three dimensional CFT with chemical potential.  The free energy is easily computed as $G = M - Ts$,
\begin{align}
 G &= \frac{1}{8 \pi } \bigg[\frac{r_+^3}{\ell^2} + \frac{q^2}{r_+} - 2 \pi r_+^2 T  +  r_+^3 \sum_{n=3}^\infty \frac{\gamma_n }{n(n-1)}   \left(\frac{4 \pi T }{r_+} \right)^{n}   \bigg] \, .
\end{align}

The equation of state can be found by rearranging the second of the two near horizon equations presented above yielding,
\be 
P = \frac{T}{v} + \sum_{n=3}^\infty  \gamma_n  \left[ \frac{2^{3n} (n-3)}{n(n-1)} \right] \left(\frac{ T }{v} \right)^n + \frac{e^2}{v^4}
\ee
where we have defined,
\be 
v = 2 r_+ \, , \quad q^2 = \frac{\pi}{2} e^2 \, ,
\ee
as the specific volume and rescaled charge parameter. 
We want to determine if it is possible for this equation of state to admit critical points. To do so requires simultaneously solving,
\be 
\frac{\partial P}{\partial v} = \frac{\partial^2 P}{\partial v^2} = 0 \, 
\ee
for temperature and volume.  

It is straightforward to see that a critical point will not result from the inclusion of a single one of the higher order terms. Assuming the $n^{th}$ order term is the only non-zero term, the two equations above can be written as,
\begin{align}
0 &=  \frac{T}{v^2}  + \gamma_n  \left[ \frac{2^{3n} (n-3)}{n-1} \right]\frac{1}{v}\left(\frac{ T }{v} \right)^n  + 4 \frac{e^2}{v^5} \, ,  
\nn\\
0 &=  2\frac{T}{v^3} +  \gamma_n  \left[ \frac{2^{3n} (n-3)(n+1)}{n-1} \right]\frac{1}{v^2 }\left(\frac{ T }{v} \right)^n + 20 \frac{e^2}{v^6} \, ,  
\end{align}
letting $x = T/v$, the first of these equations can be re-arranged for $v$:
\be 
v^4 = -\frac{4 e^2}{x + \gamma_n  \left[ \frac{2^{3n} (n-3)}{n-1} \right] x^n}
\ee
In order for $v$ to be positive, we must then have
\be 
x + \gamma_n  \left[ \frac{2^{3n} (n-3)}{n-1} \right] x^n < 0 \, .
\ee
Substituting the solution for $v$ into the second equation, we arrive at the following polynomial determining $x$,
\be 
0 = 3x -   \gamma_n \frac{2^{3n}(n-3)(n-4)}{n-1} x^{n} \, .
\ee
Now, in the case where $n=3$ or $n=4$, the only sensible solution is $x = 0$, which means there is no physical critical point.  For higher order terms ($n \ge 5$), we can substitue the solution of this equation into the previous inequality and find that it would hold provided,
\be 
1 + \frac{3}{n-4} < 0 
\ee
however, this can never be the case for $n > 4$, and therefore  there are no critical points when only a single of the higher curvature terms is present.

When more than one of the higher curvature couplings are non-zero, it is much harder to make general statements about the solutions as doing so requires drawing conclusions on the roots of polynomials of arbitrary order.  Nonetheless, it is easy to show that critical points generically occur, and we will focus here on a particular example rather than the general case.  It turns out that this example is sufficient to reveal the novel criticality that occurs due to higher order terms in four dimensions.

Consider the case where 
\be 
\gamma_3 = 0 \, , \quad \gamma_n = 0 \,\,\, \forall n > 5 \, .
\ee
In this instance, there is a single critical point, characterized by mean field theory critical exponents, and given by the critical values,
\begin{align}
T_c &= \frac{\sqrt{|e|}}{4} \left[{\frac {-6 \gamma_4^{3}-24 \sqrt {6}\gamma_4\gamma_5
^{3/2}+12 \sqrt {\gamma_5^{3/2} \left( 48\,\gamma_4^{2}\gamma_5^{3
/2}+ \sqrt {6} \left( \gamma_4^{4}+96\,\gamma_5^{3} \right)  \right) 
}}{{\pi }^{3} \left( \gamma_4^{4}-96\,\gamma_5^{3} \right) }}\right]^{1/4} \, , 
\nn\\
v_c &= 1024\,{\frac {{\pi }^{4}\gamma_5 \left( \gamma_4^{4}-96\,\gamma_5^{3
} \right) }{6\,\sqrt {6}\gamma_5^{3/2}\gamma_4^{2}+144\,\gamma_5^{
3}-3g_{
{4}} \sqrt {\gamma_5^{3/2} \left( 48\,\gamma_4^{2}\gamma_5^{3/2}+
 \sqrt {6} \left( \gamma_4^{4}+96\,\gamma_5^{3} \right)  \right) }}} \frac{T_c^5}{e^2} \, ,
 \nn\\
P_c &= \frac{3}{5}\frac{T_c}{v_c} \, . 
\end{align}
Note that the critical ratio,
\be 
\frac{P_c v_c}{T_c} = \frac{3}{5}
\ee
is distinct from what is found for spherical, charged black holes in four dimensions, where this ratio has been found to be $3/8$~\cite{Kubiznak:2012wp, Hennigar:2016gkm}, suggesting, again, a possible connection between the horizon topology and the value of this universal ratio.   The critical points are physical, i.e. respect all physicality conditions, provided the couplings satisfy,
\be 
\gamma_4 < 0 \, , \quad \text{and } \quad  0 < \gamma_5 < \left(\frac{\gamma_4^{4}}{3 \times 2^{5} } \right)^{1/3}  \, .
\ee

\begin{figure}[H]
\centering
\includegraphics[width=0.75\textwidth]{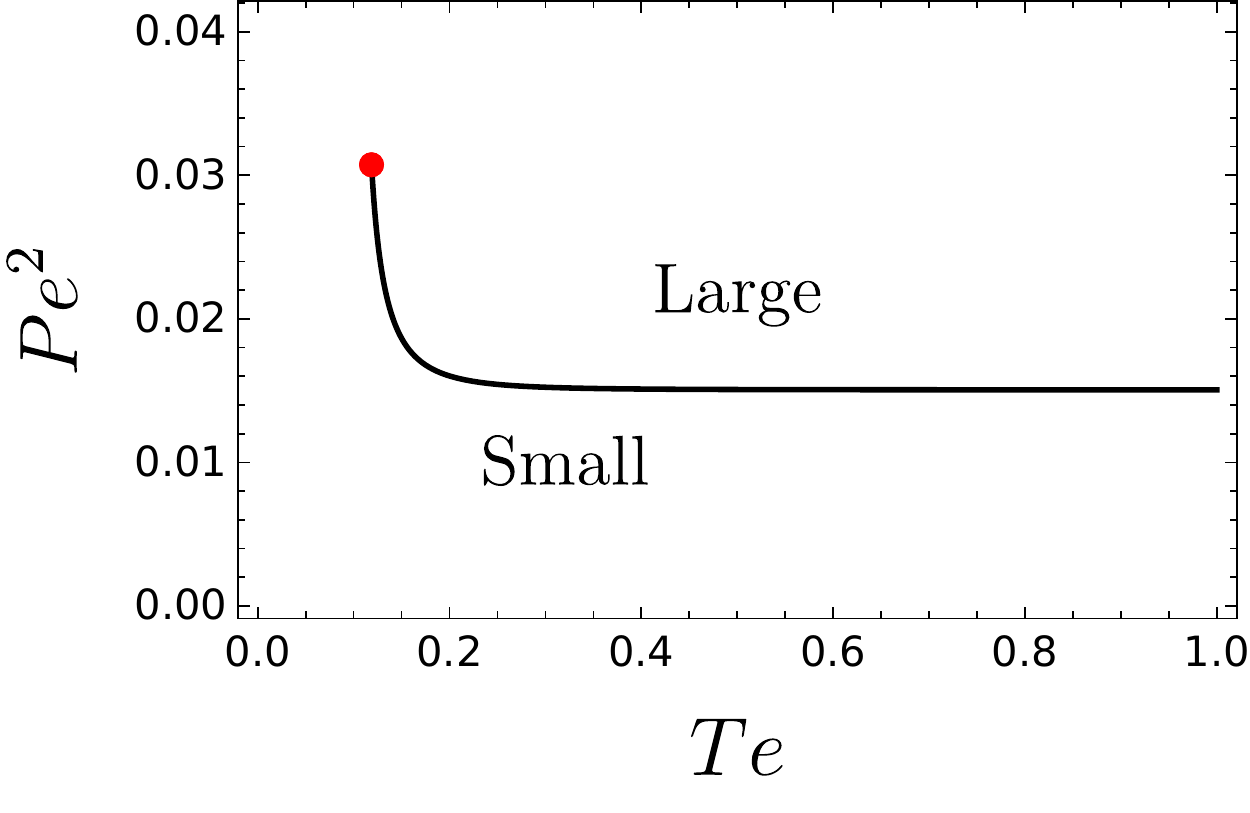}
\caption{{\bf Phase diagram in four dimensions} (color online).  The solid black line is the coexistence curve of a first order phase transition, and the red dot corresponds to the critical point.  This plot was constructed for the particular choices, $\gamma_3 = 0$, $\gamma_4 \approx -9.43$, $\gamma_5 \approx 2.18$ and $\gamma_n = 0$ for $n > 5$.  The diagram is qualitatively similar for all valid choices of the parameters.  Here, the dimensionful quantities were expressed in terms of the electric charge, which was then set to unity.   }
\label{fig:4d_phase}
\end{figure}

The phase diagram for these critical points is typical of what was observed earlier in this work, and is shown in figure~\ref{fig:4d_phase}.  The structure of the Gibbs free energy is qualitatively similar to that shown in figure~\ref{fig:5d_free_energy}.  These results are characteristic of what occurs when arbitrary higher order terms are included, and therefore we do not explicate on the higher order solutions in any more detail here.

\section{Conclusions}

We have performed a detailed study of electrically charged black branes solutions of generalized quasi-topological gravity.  Due to the lack of an analytic solution, we have constructed series solutions asymptotically, near the horizon, and near the origin.  The series approximations can be joined together numerically in the intermediate regimes.  An interesting feature we have observed is that, while the black branes possess a curvature singularity at the origin, the metric function approaches a finite, constant value as $r \to 0$, indicating that the curvature singularity is much softer.

In a consideration of the thermodynamic properties of the black branes, we  have found that despite higher derivative equations of motion, the solutions are characterized only by their mass and do not possess any higher derivative hair.  Furthermore, despite the lack of an analytic solution, the thermodynamic quantities can be obtained exactly, meaning no approximations are necessary in studying their properties.  We have also restricted attention to choices of couplings where the solutions are physically reasonable, i.e. there are no ghosts and the entropy is positive.  

We have found that these higher curvature terms generically induce critical behaviour in all dimensions $ d \ge 4$, which is a result not previously observed for charged black branes. The phase transitions occur for significant portions of the parameter space described by the higher curvature couplings. The first order phase transition begins at a critical point characterized by mean field theory critical exponents, and continues to arbitrarily high temperatures.  The structure of the free energy was found to be given by a swallowtail structure, however, instead of the swallowtail portion of the curve corresponding to a region of negative specific heat, the higher curvature terms result in positive specific heat.  Thus, there are three branches of black branes with positive specific heat, but only one is energetically favored.  This, combined with recent studies of asymptotically flat black holes~\cite{Bueno:2017qce}, suggest that it is in fact quite a general effect of the generalized quasi-topological class of theories that an increase in thermodynamic stability of black objects occurs.   

There remain a number  of interesting questions to address in regards  to these  theories.  First, in light of the results presented here, it would be worthwhile to further study the holographic aspects of the theory.  Further elaborating on the consequences of these phase transitions for the dual CFT, and also the holographic hydrodynamics would be particularly worthwhile.  The modifications to the entropy density should lead to new behaviour and perhaps interesting modifications to the ratio of shear viscosity to entropy density.  Additionally, holographic studies would allow for additional stability constraints to be placed on the higher curvature couplings.  Also, recent methods employed to study the stability of black objects in Lovelock theory could be adapted to the study this theory~\cite{Konoplya:2017lhs}.  On the classical gravity front, it would be useful to know if Birkhoff's theorem holds for the generalized quasi-topological terms, similar to Lovelock and quasi-topological gravities~\cite{Oliva:2011xu, Oliva:2012zs, Ray:2015ava}.  Furthermore, since this class of theories provide the only examples of ghost-free higher curvature theories that are non-trivial in four dimensions, it would be useful to work out any implications they have for four dimensional gravitational physics.  We will address some of these questions in future work.

\section*{Acknowledgments}
I am grateful to Robert Mann, Mozhgan Mir and  Alexander Smith for comments on an earlier version of this manuscript and to Aida Ahmadzadegan for encouragement at the early stages of this work.
This work was supported in part by the Natural Sciences and Engineering Research Council of Canada. 

\appendix

\section{Explicit forms of Lagrangian densities}
\label{sec:lagrangians}
Here we will list the various Lagrangian densities which appear in the action and their corresponding contribution to the field equations.  The $k^{th}$ order Lovelock terms are given by,
\be 
\mathcal{X}_{2k} = \frac{1}{2^k} \delta_{a_1b_1 \dots a_k b_k}^{c_1 d_1 \dots c_k d_k} R_{c_1 d_1}{}^{a_1 b_1} \cdots R_{c_k d_k}{}^{a_k b_k}
\ee
where $\delta_{a_1b_1 \dots a_k b_k}^{c_1 d_1 \dots c_k d_k} $ is totally antisymmetric in both sets of indices.  

The cubic quasi-topological Lagrangian density takes the form~\cite{Myers:2010ru, Oliva:2010eb},
\ba
\mathcal{Z}_d &=&    {{{R_a}^b}_c{}^d} {{{R_b}^e}_d}^f {{{R_e}^a}_f}^c
               + \frac{1}{(2d - 3)(d - 4)} \Bigl( \frac{3(3d - 8)}{8} R_{a b c d} R^{a b c d} R  - \frac{3(3d-4)}{2} {R_a}^c {R_c}^a R \nonumber \\
              &&  - 3(d-2) R_{a c b d} {R^{a c b}}_e R^{d e} + 3d  R_{a c b d} R^{a b} R^{c d}
                + 6(d-2) {R_a}^c {R_c}^b {R_b}^a  + \frac{3d}{8} R^3 \Bigr)\,.\quad
\label{Zd}
\ea
and the generalized quasi-topological term is given by, 
\ba\label{Sd}
\mathcal{S}_{3, d} &=&
14 R_{a}{}^{e}{}_{c}{}^{f} R^{abcd} R_{bedf}+ 2 R^{ab} R_{a}{}^{cde} R_{bcde}- \frac{4 (66 - 35 d + 2 d^2) }{3 (d-2) (2 d-1)} R_{a}{}^{c} R^{ab} R_{bc}\nonumber\\
&& -  \frac{2 (-30 + 9 d + 4 d^2) }{(d-2) (2 d-1)} R^{ab} R^{cd} R_{acbd} -  \frac{(38 - 29 d + 4 d^2)}{4 (d -2) (2 d  - 1)} R R_{abcd} R^{abcd}  \nonumber\\
&&+ \frac{(34 - 21 d + 4 d^2) }{(d-2) ( 2 d - 1)} R_{ab} R^{ab} R -  \frac{(30 - 13 d + 4 d^2)}{12 (d-2) (2 d - 1)}  R^3 
\ea
which was defined in~\cite{Hennigar:2017ego}.  

At the quartic level, there are four generalized quasi-topological terms in dimensions larger than four and six in four dimensions.  However, in each case, the field equations are identical and we can, without loss of generality, chose any representative of this class.  Here we have chosen the following expression,

\begin{align}
\mathcal{S}_{4, d} &= \frac{1}{3 (d - 3)^2 (d - 2) (d -1) d (11 - 6 d + d^2) (19 - 18 d + 3 d^2) (-22 + 26 d - 9 d^2 + d^3)} \times
\nn\\
&\times \big[ 
-4 (d - 2) (-718080 + 2405582 d - 3666144 d^2 + 3359133 d^3 - 2057938 d^4 
\nn\\
&+ 887142 d^5 - 276120 d^6 + 62662 d^7 - 10296 d^8 + 1182 d^9 - 86 d^{10} + 3 d^{11}) R_{a}{}^{c} R^{ab} R_{b}{}^{d} R_{cd} 
\nn\\
&- 4 (707880 - 2115012 d + 2700668 d^2 - 1809780 d^3 + 561468 d^4 + 61133 d^5 - 134394 d^6 
\nn\\
&+ 60426 d^7 - 15005 d^8 + 2238 d^9 - 189 d^{10} + 7 d^{11}) R_{ab} R^{ab} R_{cd} R^{cd} 
\nn\\
&+ 16 (d - 2) (d -1) (-8670 + 30262 d - 47247 d^2 + 43299 d^3 - 25747 d^4 + 10271 d^5 
\nn\\
&- 2734 d^6 + 466 d^7 - 46 d^8 + 2 d^9) R_{a}{}^{c} R^{ab} R_{bc} R + 4 (198900 - 592178 d + 790224 d^2 
\nn\\
&- 617415 d^3 + 313537 d^4 - 109500 d^5 + 27237 d^6 - 4900 d^7 + 624 d^8 - 51 d^9 + 2 d^{10}) R_{ab} R^{ab} R^2 
\nn\\
&+ 48 (d - 2) (-68340 + 203532 d - 268574 d^2 + 203038 d^3 - 95967 d^4 + 29190 d^5 
\nn\\
&- 5665 d^6 + 667 d^7 - 42 d^8 + d^9) R^{ab} R^{cd} R R_{acbd} - 6 (192100 - 603774 d + 820554 d^2 
\nn\\
&- 605255 d^3 + 237492 d^4 - 22951 d^5 - 24843 d^6 + 14329 d^7 - 3890 d^8 + 609 d^9 - 53 d^{10} 
\nn\\
&+ 2 d^{11}) R^2 R_{abcd} R^{abcd} - 48 (d - 2) (d -1) (-63580 + 183572 d - 244118 d^2 
\nn\\
&+ 192444 d^3 - 97734 d^4 + 32893 d^5 - 7308 d^6 + 1032 d^7 - 84 d^8 + 3 d^9) R_{a}{}^{c} R^{ab} R^{de} R_{bdce}
\nn\\
& + 12 (d - 2) (d -1) (-29920 + 120000 d - 196892 d^2 + 175930 d^3 - 93864 d^4 
\nn\\
&+ 30115 d^5 - 5212 d^6 + 193 d^7 + 99 d^8 - 18 d^9 + d^{10}) R^{ab} R^{cd} R_{ac}{}^{ef} R_{bdef} 
\nn\\
&+ 4 (d - 3) (d - 2) (d -1) (2550 - 15414 d + 28633 d^2 - 26167 d^3 + 13715 d^4 
\nn\\
&- 4351 d^5 + 830 d^6 - 88 d^7 + 4 d^8) R R_{ab}{}^{ef} R^{abcd} R_{cdef} + 6 (-60520 + 414664 d 
\nn\\
&- 945458 d^2 + 1097752 d^3 - 719367 d^4 + 242784 d^5 - 3125 d^6 - 36155 d^7 + 17569 d^8 
\nn\\
&- 4430 d^9 + 659 d^{10} - 55 d^{11} + 2 d^{12}) R_{ab} R^{ab} R_{cdef} R^{cdef}
\nn\\
& - 12 (d - 3) (d - 2) (d -1) (-22 + 26 d - 9 d^2 + d^3) (-340 + 494 d - 70 d^2 - 185 d^3 
\nn\\
&+ 112 d^4 - 25 d^5 + 2 d^6) R^{ab} R_{a}{}^{cde} R_{bc}{}^{fh} R_{defh} \big]
\nn\\
&+  R_{ab}{}^{ef} R^{abcd} R_{ce}{}^{hj} R_{dfhj}
\end{align}
which corresponds to $\mathcal{S}_{d}^{(3)}$ in~\cite{Ahmed:2017jod}.  

\bibliography{mybib,LBIB}
\bibliographystyle{JHEP}
\end{document}